 \documentclass[acmsmall]{acmart}
\AtBeginDocument{%
  \providecommand\BibTeX{{%
    \normalfont B\kern-0.5em{\scshape i\kern-0.25em b}\kern-0.8em\TeX}}}

\usepackage{algorithmic}
\usepackage{graphicx}
\usepackage{textcomp}
\usepackage{xcolor}
\usepackage{multirow,color, hyperref, url, graphics}
\usepackage{listings}
\usepackage{diagbox}
\usepackage{subfig}
\usepackage{balance}
\usepackage{adjustbox}
\usepackage{colortbl}
\usepackage{framed}
\usepackage{booktabs}
\usepackage{indentfirst}
\usepackage{makecell,tcolorbox}
\usepackage[framemethod=TikZ]{mdframed}

\def\renderinsertedpart{}

\ifx\renderinsertedpart\undefined

\newenvironment{insertedpartenv}{}{}
\else

\fi

\newcommand{\finding}[1]{
  \vspace{1.5mm}
 \begin{mdframed}[linecolor=gray,roundcorner=12pt,backgroundcolor=gray!15,linewidth=3pt,innerleftmargin=2pt, leftmargin=0cm,rightmargin=0cm,topline=false,bottomline=false,rightline = false]
  #1
 \end{mdframed}
 \vspace{1.5mm}
}

\begin{document}

% \title{Knowledge-based GUI Test Case Generation}
\title{GUI Test Migration via Abstraction and Concretization}

\author{Yakun Zhang}
\email{zhangyakun@stu.pku.edu.cn}
\affiliation{%
  \institution{Key Lab of HCST (PKU), MOE; SCS\\Peking University}
  \city{Beijing}
  \country{China}}

\author{Chen Liu}
\email{cissieliu@stu.pku.edu.cn}
\affiliation{%
  \institution{Key Lab of HCST (PKU), MOE; SCS\\Peking University}
  \city{Beijing}
  \country{China}}

\author{Xiaofei Xie}
\email{xfxie@smu.edu.sg}
\affiliation{%
  \institution{Singapore Management University}
  \city{Singapore}
  \country{Singapore}}

\author{Yun Lin}
\email{lin\_yun@sjtu.edu.cn}
\affiliation{%
  \institution{Shanghai Jiao Tong University}
  \city{Shanghai}
  \country{China}}

\author{Jin Song Dong}
\email{dcsdjs@nus.edu.sg}
\affiliation{%
  \institution{National University of Singapore}
  \city{Singapore}
  \country{Singapore}}

\author{Dan Hao}
\email{haodan@pku.edu.cn}
\affiliation{%
  \institution{Key Lab of HCST (PKU), MOE; SCS\\Peking University}
  \city{Beijing}
  \country{China}}

\author{Lu Zhang}
\authornote{Corresponding author}
\email{zhanglucs@pku.edu.cn}
\affiliation{%
  \institution{Key Lab of HCST (PKU), MOE; SCS\\Peking University}
  \city{Beijing}
  \country{China}}

% \author{Lu Zhang}
% \affiliation{%
%   \institution{Peking University}
%   \city{Beijing}
% \email{zhanglucs@pku.edu.cn}

\renewcommand{\shortauthors}{Zhang et al.}

\begin{abstract}
GUI test migration aims to produce test cases with events and assertions to test specific functionalities of a target app. Existing migration approaches typically focus on the widget-mapping paradigm that maps widgets from source apps to target apps. However, since different apps may implement the same functionality in different ways, direct mapping may result in incomplete or buggy test cases, thus significantly impacting the effectiveness of testing the target functionality and the practical applicability of migration approaches.

In this paper, we propose a new migration paradigm (i.e., the abstraction-concretization paradigm) that first abstracts the test logic for the target functionality and then utilizes this logic to generate the concrete GUI test case. Furthermore, we introduce \emph{\toolNameSmall{}}, the first approach that migrates GUI test cases based on this paradigm.
Specifically, we propose an abstraction technique that utilizes source test cases from source apps targeting the same functionality to extract a general test logic for that functionality.
Then, we propose a concretization technique that utilizes the general test logic to guide an LLM in generating the corresponding GUI test case (including events and assertions) for the target app.
We evaluate \toolNameSmall{} on two widely-used datasets (including 31 apps, 34 functionalities, and 123 test cases). On the FrUITeR dataset, the test cases generated by \toolNameSmall{} successfully test 64\% of the target functionalities, improving the baselines by 191\%.
On the Lin dataset, \toolNameSmall{} successfully tests 75\% of the target functionalities, outperforming the baselines by 42\%.
These results underscore the effectiveness of \toolNameSmall{} in GUI test migration.
\end{abstract}

\begin{CCSXML}
  <ccs2012>
  <concept>
  <concept_id>10011007.10011074.10011099.10011102.10011103</concept_id>
  <concept_desc>Software and its engineering~Software testing and debugging</concept_desc>
  <concept_significance>300</concept_significance>
  </concept>
  </ccs2012>
\end{CCSXML}

\ccsdesc[300]{Software and its engineering~Software testing and debugging}

\acmJournal{TOSEM}
\acmVolume{37}
\acmNumber{4}
\acmArticle{111}
\acmMonth{12}
\acmDOI{10.1145/XXXXXXX.XXXXXXX}

\keywords{Test migration, Functional GUI testing, Large language model}



\graphicspath{{figures/}}
\newcommand{\figref}[1]{Figure~\ref{#1}}
\newcommand{\tabref}[1]{Table~\ref{#1}}
\newcommand{\secref}[1]{Section~\ref{#1}}

\newcommand{\toolName}[1]{\textbf{MACdroid\raisebox{-0.7ex}{\Huge\textbf{#1}}}}

\newcommand{\toolNameplain}[1]{MACdroid\raisebox{-0.7ex}{\Huge #1}}

\newcommand{\toolNameSmall}{MACdroid}

\maketitle
\newcommand{\yakun}[1]{\textcolor{blue}{[yakun: #1]}}

\newcommand{\ie}{\mbox{\emph{i. e.,\ }}}

\section{Introduction}\label{sec:introduction}
Graphical User Interface (GUI) testing is common for testing functionalities of mobile apps~\cite{baek2016automated,us2018functionality,liu2023legodroid}. A GUI test case is composed of some ordered \emph{events} and \emph{assertions}~\cite{behrang2019test,lin2019test}. 
These events are designed to probe the functionalities of GUI \emph{widgets}. 
These assertions verify whether the outcomes of the events align with developers' expectations. 
Developers typically develop multiple functionalities (e.g., sign-in) within an app. The target functionality of a GUI test case refers to
the specific functionality that the test case is designed to verify.
Automatic generation of GUI test cases is challenging due to the limited understanding of the specific functionality in target apps.
Consequently, GUI test cases are still predominantly crafted manually, which is time-consuming and labor-intensive~\cite{dobslaw2019estimating,pan2020reinforcement,linares2017developers,bai2023argusdroid}.
To reduce the manual effort in writing GUI test cases, several migration approaches~\cite{zhang2024learning, lin2019test, behrang2019test, mariani2021semantic} have been proposed. These approaches migrate GUI test cases from a source app to a target app by mapping widgets that are semantically similar.

Despite advancements in migration approaches, the migrated test cases often remain incomplete or contain bugs~\cite{zhao2020fruiter}, making them challenging to directly use in real-world scenarios. This issue arises because existing migration approaches follow the \emph{widget-mapping paradigm}, which involves mapping widgets from source test cases (i.e., test cases for source apps) to target test cases (i.e., test cases for target apps). However, different apps may implement the same functionality in different ways. Source widgets (i.e., widgets in the source apps) may not be similar to target widgets (i.e., widgets in the target apps). As a result, test cases generated based on the widget-mapping paradigm might lack some necessary widgets, leading to an incomplete test of the target functionality. For example, the sign-in functionality in some apps might require an email and a password, while other apps might require a phone number and a password. 
Directly migrating individual test cases may not completely test the target functionality of the target app.

Considering the significant disparity between test cases generated by existing migration approaches~\cite{zhang2024learning, lin2019test, behrang2019test, mariani2021semantic} and their target functionalities, it is crucial to propose a new migration paradigm capable of generating high-quality GUI test cases. Inspired by existing research~\cite{lin2019test, behrang2019test}, we observe that although different apps may have variations in their specific implementation of the same functionality, the underlying logics for the target functionality are typically similar.
In other words, test cases targeting the same functionality tend to follow similar test logics. 
For instance, the typical test logic for the sign-in functionality of a shopping app involves navigating to the sign-in state, completing all the required fields, clicking the sign-in related button, and verifying the correct display of user information.
This test logic abstracts away the implementation details of specific mobile apps (e.g., using email or phone number) while preserving the core concept of the testing process, making it an effective guide for generating test cases related to the target functionality.

Based on the preceding observation, we propose a new migration paradigm, the \emph{abstraction-concretization paradigm}, which first abstracts a general test logic of the target functionality from multiple source test cases, and then uses this logic to guide the generation of concrete GUI test case for the target app.
In the abstraction phase, we eliminate the app-specific details, focusing solely on the general test logic of the target functionality. In the concretization phase, we apply the general test logic to generate the concrete target events and assertions.
Additionally, the emergence of large language models~\cite{chatgpt,gpt-4,claude}(LLMs) brings new opportunities for GUI test migration. With sophisticated semantic understanding and reasoning capabilities~\cite{chen2025deep}, LLMs have the potential to understand target functionality and tackle the complexities associated with test logic abstraction and event/assertion generation, thereby enhancing the effectiveness of GUI test migration.

Specifically, we introduce a two-stage approach named \toolNameSmall{} (i.e., \textbf{M}igrating GUI test cases via \textbf{A}bstraction and \textbf{C}oncretization). 
First, we propose an \textbf{abstraction technique} that utilizes source test cases to extract a \emph{general test logic}. Initially, we extract an individual test logic as a sequence of structured test steps for each source test case. These test steps retain only functionality-related information, facilitating adaptation to new apps. Subsequently, we integrate the individual test logics from multiple source test cases into a sequence of structured test steps, forming a general test logic. The general test logic summarized from multiple test cases provides a comprehensive perspective on the target functionality, facilitating complete testing of the target functionality.
Second, we propose a \textbf{concretization technique} that utilizes the extracted general test logic to guide an LLM in the step-by-step concretization of the target test case. Initially, we identify a set of privileged events and assertions derived from source test cases, which may be relevant to testing the functionality of the target app. To improve accuracy, we design a priority strategy that selects events and assertions from this privileged set, rather than directly selecting from all operable widgets in the target app. This restriction narrows the candidate set, enhancing the accuracy of event and assertion selection.  Moreover, we design a validation mechanism that identifies and repairs potential inaccuracies by comparing the general test logic and the output of the LLM, ensuring the accuracy and reliability of the generated test case.

We conduct a comprehensive evaluation to analyze the effectiveness of \toolNameSmall{} using 31 real-world apps, 34 functionalities, and 123 test cases from the FrUITeR dataset~\cite{fruiter-dataset} and the Lin dataset~\cite{craftdroid-dataset}. 
We compare \toolNameSmall{} with the state-of-the-art migration approach TEMdroid~\cite{zhang2024learning} and the state-of-the-art generation approach AutoDroid~\cite{wen2023empowering} on these datasets. On the FrUITeR dataset, the test cases generated by \toolNameSmall{} successfully test 64\% of the target functionalities, representing a 191\% improvement over the baselines. On the Lin dataset, \toolNameSmall{} successfully tests 75\% of the target functionalities, outperforming the baselines by 42\%. 
We also evaluate the effectiveness of \toolNameSmall{}'s main techniques for the GUI test migration.
Overall, these results demonstrate that \toolNameSmall{} is effective in GUI test migration for industrial apps.

This paper makes the following main contributions:
\begin{itemize}
    \item We propose a new migration paradigm (i.e., the abstraction-concretization paradigm), and introduce \toolNameSmall{}, the first approach that follows this paradigm to migrate GUI test cases.
    \item We propose a novel technique for automatically abstracting a general test logic from source test cases for the target functionality.
    \item We propose a novel technique for concretizing GUI test cases that focuses on accurately selecting events and assertions for the target apps.
    \item We conduct an empirical evaluation using real-world apps, demonstrating the effectiveness of \toolNameSmall{}. The source code of \toolNameSmall{} is publicly available~\cite{ftdroid}.
\end{itemize}

\section{Illustrative Example}\label{sec:example}
\begin{figure*}[htbp]
	\center
 \includegraphics[scale = 0.25]{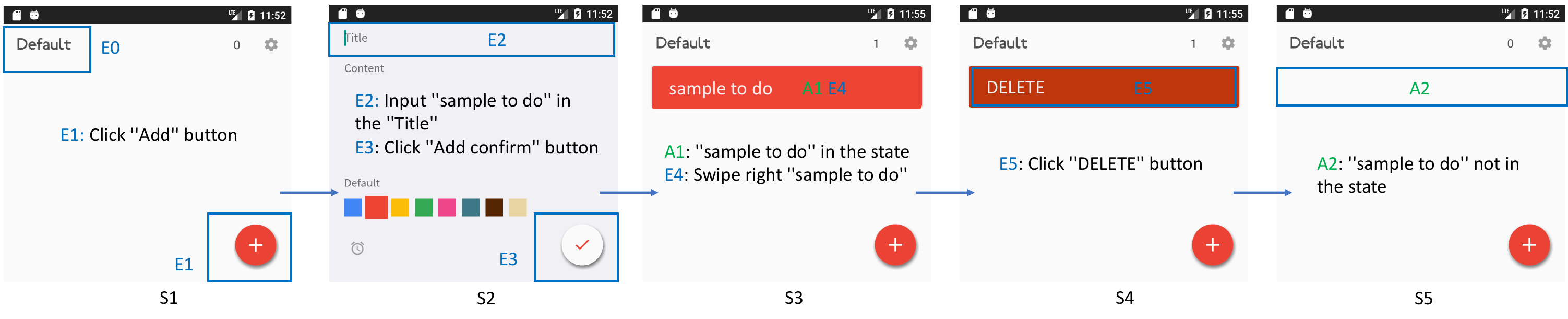}
	\caption{The test process of ``Add and remove an item'' in a To-do app}
	\label{fig:motivation-example}
\end{figure*}

\figref{fig:motivation-example} depicts the test process for the functionality of ``Add and remove an item'' in a To-do app based on the test case from the Lin dataset~\cite{craftdroid-dataset}. 
This test case is designed to validate whether a user can successfully add a to-do item in the target app and subsequently remove it after finishing this item. 
We use this example to illustrate the motivation of this paper. We also use this example to illustrate the methodology of \toolNameSmall{} in the following sections.

Specifically, associated with five GUI states (i.e., S1 to S5), the test case includes five events (i.e., E1 to E5) and two assertions (i.e., A1 and A2). The test process is that a user clicks the ``Add'' button (E1),  inputs ``sample to do'' in the ``Title'' box (E2), and clicks the ``Add confirm'' button (E3). One assertion (A1) checks for successfully adding one item by verifying that ``sample to do'' appears in the new state (S3). The user then swipes right on the ``sample to do'' item (E4) and clicks the ``DELETE'' button (E5). The other assertion (A2) checks for successfully removing this item by verifying that ``sample to do'' no longer appears in the new state (S5).

We evaluate the effectiveness of existing approaches on the preceding example. Specifically, we select the state-of-the-art test case migration approach, TEMdroid~\cite{zhang2024learning}, and the state-of-the-art functional test generation approach, AutoDroid~\cite{wen2023empowering}.  However, neither approach can fully test the target functionality. This limitation arises because the specific implementations of the same functionality vary across apps, making it difficult to comprehensively test the target app’s functionality by directly migrating or generating test cases. Furthermore, incorrect migrations or generations result in erroneous events and assertions, preventing these test cases from successfully testing the target functionality.

The results obtained by existing approaches cannot be directly used in industry and still require manual modification.
Given the substantial disparity between the test cases generated by existing approaches and their corresponding target functionalities, it is imperative to propose a new migration paradigm capable of generating high-quality GUI test cases.

\section{MACdroid}\label{sec:ftdroid}

\begin{figure}[t]
	\center
 \includegraphics[width=12cm]{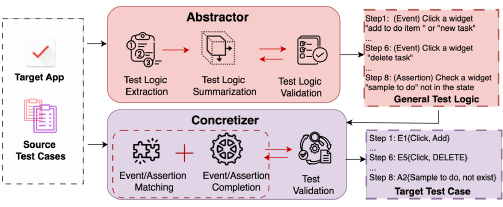}
	\caption{Overview of \toolNameSmall{}}
	\label{fig:workflow}
% \vspace{-2mm}
\end{figure}

Given a target app and a set of source test cases that test the same functionality in source apps as inputs, \toolNameSmall{} (whose workflow is shown in \figref{fig:workflow}) generates a target test case to test the target functionality based on two components.
The first component is the \textbf{Abstractor component}. This component aims to extract a general test logic based on multiple source test cases (see \secref{sec:abstractor}).
The second component is the \textbf{Concretizer component}. 
This component aims to concrete target test cases according to test generation (including Event/Assertion Matching and Event/Assertion Completion) and test validation (see \secref{sec:generator}).

%figure

% \vspace{-0.2cm}

\subsection{Abstractor}
\label{sec:abstractor}

Given the source test cases for the target functionality, the Abstractor component (see \figref{fig:workflow}) aims to extract a general Test Logic (abbr. TL) for this functionality. The variability among these source test cases presents a challenge in extracting the test logic for the target functionality. To address this challenge, Abstractor includes three modules: \emph{TL Extraction}, \emph{TL Summarization}, and \emph{TL Validation}.

In the \textbf{TL Extraction} module, \toolNameSmall{} extracts a sequence of structured test steps from each source test case, referred to as an individual test logic (see \secref{sec:test-representation}). This step-based format facilitates easier comprehension by LLMs~\cite{wei2022chain}. Note that, during this process, app-specific details are isolated to emphasize only functionality-related information, thereby enhancing adaptability across different apps.

After extraction, the individual test logic for different source test cases may vary. This is because different apps may implement the same functionality differently, the individual test logic for each source test case may be different and may only cover partial test steps for the target functionality. The \textbf{TL Summarization} module addresses this problem by creating a general and comprehensive test logic suitable for diverse apps. Since LLMs have demonstrated the capabilities to understand and summarize information, this step adopts an LLM to summarize a general and comprehensive test logic based on commonalities and differences among the individual logics (see \secref{sec:step-summarization}). 
Furthermore, recognizing potential inaccuracies in LLMs' outputs (e.g., hallucination issues~\cite{yao2023llm, martino2023knowledge} and comprehension biases~\cite{chen2024humans,han2024instinctive}), the \textbf{TL Validation} module employs a rule-based method to ensure the quality of the summarized test logic (see \secref{sec:step-validation}). Finally, Abstractor outputs a refined, general, and comprehensive test logic for the target functionality, which could be applied to various apps.

%figure
\begin{figure}[t]
	\center
 \includegraphics[scale = 0.4]{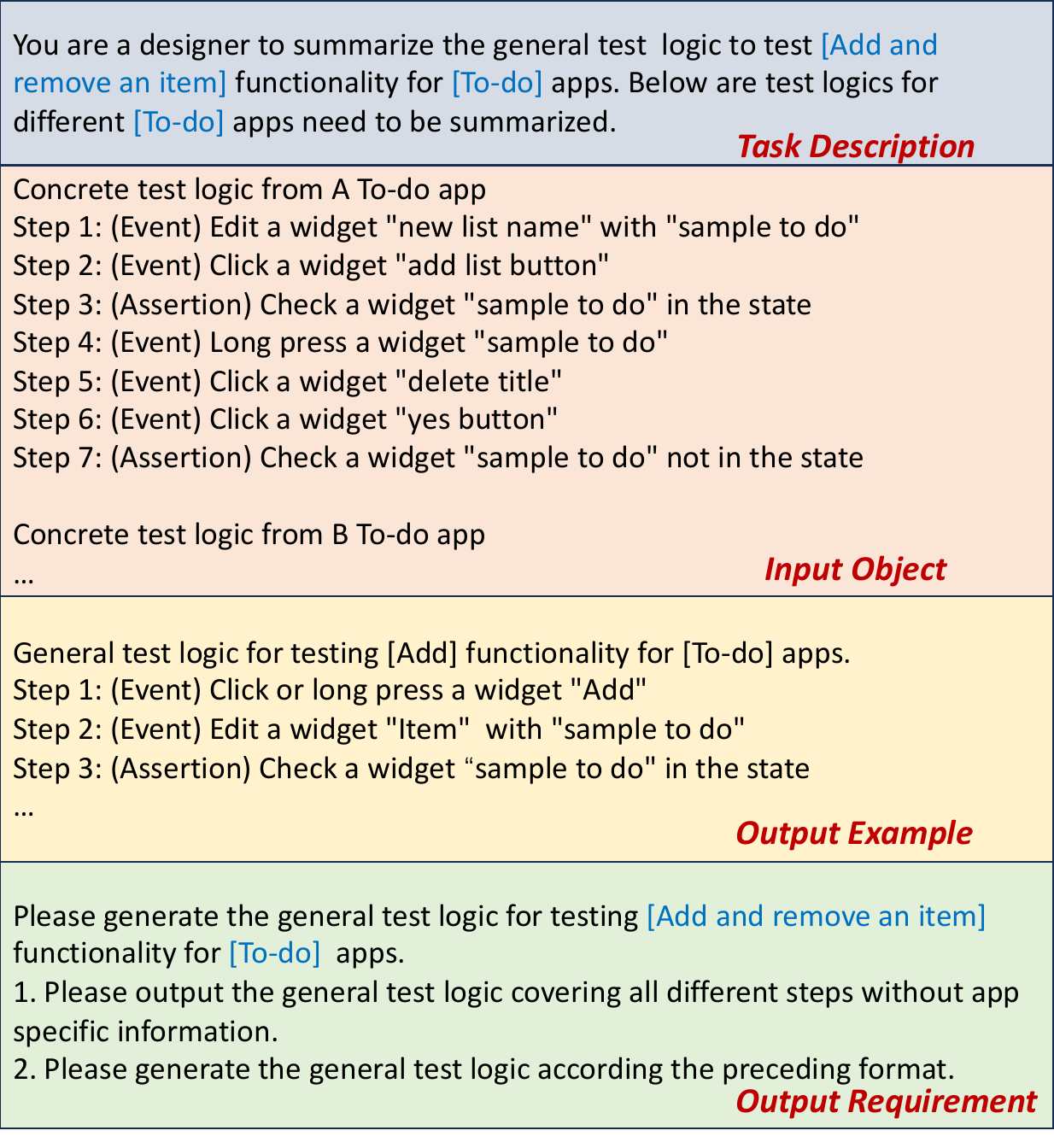}
	\caption{A prompt example used by TL Summarization}
 % \vspace{-5mm}
	\label{fig:prompt-abstractor}
\end{figure}

\subsubsection{TL Extraction}
\label{sec:test-representation}

We have two considerations for extracting the test logic from each source test case.
First, standardizing the input with a consistent format and terminology makes it easier for an LLM to understand the specific task and eliminate ambiguities arising from diverse expressions~\cite{zhang2024algo, hu2024prompting, feng2024prompting}. 
Thus, for source test cases from different test frameworks (e.g., Appium~\cite{appium} and Selenium~\cite{selenium}) and different programming languages (e.g., Python and Java), we represent the extracted test logic in a unified step-based structure.
Second, source test cases may include app-specific information (e.g., specific app environment or interaction methods). To enhance the generalizability of the extracted test logic to new apps, we extract only the essential information from the test cases and remove the app-specific information.

Based on the preceding two considerations, we represent a test logic using a sequence of ordered test steps \{$S_{e1}$, $S_{a1}$,...\}. There are two types (i.e., event and assertion) of test steps. A test step with an event type $S_e$ is represented as a tuple with four elements $(e, w, a, v)$, where $e$ represents the event type; $w$ represents a widget; $a$ represents an action to the widget; and $v$ represents an optional input value. A test step with an assertion type $S_a$ is represented as a tuple with three elements $(a, w, c)$, where $a$ represents the assertion type; $w$ represents a widget; and $c$ represents a condition to check the widget.

The ``input object'' of \figref{fig:prompt-abstractor} provides an example of using this module to extract an individual test logic from a source test case. Specifically, \toolNameSmall{} divides the source test case into a sequence of events and assertions based on keywords (e.g., ``gui'', ``assertion''). For each \textbf{event}, \toolNameSmall{} extracts the widget, the action (e.g., click, edit, swipe, scroll, and long-press), and the optional input values. A widget typically has several attributes. To represent a widget accurately and concisely, we only use the \emph{text}, \emph{content-desc}, and \emph{resource-id} attributes. After extracting these items, \toolNameSmall{} structures each event as the template ``\emph{(Event) [Action] a widget [Widget] with [Value]}'' (e.g., ``Step 1'' of ``Input Object'' in \figref{fig:prompt-abstractor}). 
For each \textbf{assertion}, \toolNameSmall{} extracts both the widget and its associated condition. After extracting these items, \toolNameSmall{} structures each assertion as the template ``\emph{(Assertion) Check a widget [Widget] [Condition]}'' (e.g., ``Step 3'' of ``Input Object'' in \figref{fig:prompt-abstractor}).

\begin{table}[htbp]
  \centering
  \caption{An introduction of prompt template used in MACdroid}
    \begin{tabular}{l|l}
    \toprule
    \multicolumn{1}{c|}{\textbf{Part}} & \multicolumn{1}{c}{\textbf{Aim}} \\
    \midrule
    Task description & Provide the overview goal of corresponding module. \\
    Input Object & Provide the input information to be processed by LLMs. \\
    Output example & Provide an example with the expected output formats. \\
    Output requirement & Provide the requirements and considerations for the output results. \\
    \bottomrule
    \end{tabular}%
  \label{tab:prompt-template}%
\end{table}%

\subsubsection{TL Summarization}
\label{sec:step-summarization}

This module aims to summarize a general test logic from the extracted individual test logics of multiple source test cases. 
Considering the powerful comprehension and summarization capabilities of LLMs, as well as their adaptability to diverse and complex real-world scenarios, we guide LLMs in summarizing test logics rather than relying on manual summaries or rule-based techniques. This approach enables LLMs to capture the general logics of testing functionalities across a broader range of scenarios with full automation, thereby eliminating the need for manual intervention. To ensure that LLMs have a deep understanding of the logic summarization task, we design a clear and structured guidance mechanism.
Specifically, we design a prompt with four parts, i.e., task description, input object, output example, and output requirement to interact with an LLM. The aims of these four parts are illustrated in \tabref{tab:prompt-template}.
\figref{fig:prompt-abstractor} is an example of the prompt input of the TL Summarization module.

\underline{Task description.} 
This part provides an overview guidance for the LLM to understand the aim of this module, which is summarizing multiple individual test logics into a general test logic. For different test cases, we only change the functionality to be tested and the category of the target app, which are highlighted in blue (see \figref{fig:prompt-abstractor}).
Note that, the functionality to be tested can be extracted from the function name of source test cases. The category of the target app can be extracted from the app location (e.g., Google Play Store~\cite{google-play} or F-Droid~\cite{fdroid}).

\underline{Input object.} This part displays the individual test logic extracted from each source test case by the TL Extraction module, which will be summarized.

\underline{Output example.} One-shot prompting, which enables LLMs to acquire task-specific input-output formats, has demonstrated superior performance compared to zero-shot setups~\cite{brown2020language, ahuja2023mega}. Thus, we use one-shot prompting in this prompt design. To illustrate various types with shorter texts, we design an example with all the different output formats.

\underline{Output requirement.}  This part outlines two output requirements that guide the LLM in effectively summarizing the general test logic.
Specifically, the first requirement aims to enhance the comprehensiveness and the generalizability of the general test logic. Note that, a single test step can be implemented in multiple ways (e.g., through different actions). For example, in different apps, deleting an item can be implemented in various actions, such as swiping or clicking the item.  In this situation, the LLM should use ``or'' to align these various implementations within one test step. 
The second requirement focuses on improving the quality of the general test logic by including all important information while remaining concise.
Specifically, this requirement ensures that the outputs of the LLM follow the event structure (i.e., ``\emph{(Event) [Action] a widget [Widget] with [Value]}'') and the assertion structure (i.e.,  ``\emph{(Assertion) Check a widget [Widget] [Condition]}'') as outlined in \secref{sec:test-representation}.
This structured template makes the general test logic semantically clear within the word limit of LLMs inputs~\cite{chatgpt,gpt-4}.

\begin{table}[t]
  \centering
  \caption{An introduction of validation rules}
        \resizebox{\textwidth}{!}{
    \begin{tabular}{l|l|l}
    \toprule
    \multicolumn{1}{c|}{\textbf{Rule}} & \multicolumn{1}{c|}{\textbf{Aim}} & \multicolumn{1}{c}{\textbf{Component}} \\
    \midrule
    Irrelevant step & Verify whether the general test logic contains any irrelevant test steps. & \multirow{3}[2]{*}{Abstractor} \\
    Missing step & Verify whether the general test logic lacks any necessary test steps. &  \\
    Ambiguous action & Verify whether the actions in the general test logic align with the required actions. &  \\
    \midrule
    Incorrect type & Verify whether the matched types align with the types of the specific test steps. & \multirow{4}[2]{*}{Concretizer} \\
    Irrelevant matching & Verify whether the matched events and assertions belong to the privileged set. &  \\
    Completion checking & Verify whether the generated events and assertions of the specific test steps are complete. &  \\
    Incorrect format & Verify whether the output test cases align with the required formats. &  \\
    \bottomrule
    \end{tabular}}%
  \label{tab:validation-rule}%
\end{table}%

\subsubsection{TL Validation}
\label{sec:step-validation}

The process of summarizing multiple individual test logics into a single general test logic involves combining test steps that achieve the similar objectives and also adding test steps related to achieve different objectives. This process may result in the general test logic being longer than a single individual test logic but not excessively long, as these individual test logics target at testing the same functionality and share several common test steps.
However, due to hallucination issues of LLMs, outputs of LLMs may be irrelevant or fabricated with the input data~\cite{yao2023llm, martino2023knowledge} even if the output requirements are included.
To address these potential issues and ensure the quality of the general test logic, we design three rules to identify common issues related to the TL Summarization module. Subsequently, for each issue, we generate corresponding feedback for the TL Summarization module to re-summarize the general test logic accordingly. The first three rows of \tabref{tab:validation-rule} provide an overview of the validation rules.

\underline{Irrelevant step.} 
The general test logic generated by the TL Summarization module may not only include a summary of the test steps from the source test cases but also create irrelevant test steps. We use the longest source test case as a \emph{reference}, and use $Max\_ratio$ to measure whether the general test logic remains within a reasonable length. If the generated test logic is too long (i.e., too many test steps), it may contain redundancy and irrelevant steps.

Specifically, \toolNameSmall{} calculates the ratio of the number of test steps in the general test logic to the number of test steps in the longest source test case. If this ratio exceeds $Max\_ratio$, irrelevant steps may be introduced.
To repair this issue, we design a feedback prompt as ``\emph{The number of your summarized test steps is more than the maximum number of test steps, which may introduce irrelevant test steps. Please re-summarize it}''. This feedback prompt is then sent to the TL Summarization module for re-summarization.

\underline{Missing step.} 
The general test logic generated by the TL Summarization module may lack necessary test steps, resulting in an incomplete summary.
We use the shortest test case as a \emph{reference} for validation. This is because general test logic is derived by summarizing the individual test logics from multiple test cases. The process involves merging similar steps and supplementing distinct ones. Therefore, it is rare for the general test logic to be shorter than the corresponding shortest test case.

Specifically, \toolNameSmall{} compares the number of test steps in the general test logic. If the generated general test logic is shorter than the shortest test case, that means the general test logic may lack some necessary test steps. To repair this issue, we design a feedback prompt to the TL Summarization module  for re-summarization.  The feedback prompt is ``\emph{The number of your summarized test steps is less than the minimum number of test steps, which may miss some necessary test steps. Please re-summarize it.}''.

\underline{Ambiguous action.} The actions specified in the general test logic generated by the TL Summarization module may differ from the expected actions, increasing the difficulty for the LLM to understand the functionality in the subsequent stage. 
For instance, the click action can be described in various ways, such as ``tap a screen'' or ``touch by finger'', all of which are same semantics. However, these diverse descriptions increase the difficulty for LLMs. To mitigate this issue, we standardize the description of the click action to consistently use ``click'', as outlined in the ``output example'' (see \figref{fig:prompt-abstractor}). This standardization reduces ambiguity, and enhances the accuracy and consistency of the LLM's understanding.
To identify this issue, \toolNameSmall{} compares the actions generated by the TL Summarization module with the expected actions outlined in the ``output example''. Specifically, for events, the expected actions include ``click'', ``edit'', ``swipe'', ``scroll'', and ``long-press''. For assertion, the expected action is ``check''.
If \toolNameSmall{} identifies discrepancies between the generated actions and the expected actions, it then generates a corresponding feedback prompt to the TL Summarization module. The feedback prompt is as follows: ``\emph{The [Step] does not include an action that appears in the output example. Please select one action in the output example to re-describe this step}''.

% \vspace{-0.2cm}
\subsection{Concretizer}
\label{sec:generator}

The Concretizer component aims to generate a test case for the specified functionality of the target app, utilizing the general test logic and the source test cases. The test logic summarized by the Abstractor component is general, which cannot be directly utilized as a test case for the target app. 
To concretize an executable test case, \toolNameSmall{} needs to concretize each test step in the general test logic with the specific events and assertions in the target app. 
However, it is challenging to select appropriate events and assertions due to the large number of candidates in the target app. For example, the BBC News~\cite{BBC} app includes more than 30 GUI states, and each state has an average of 60 operable widgets.

To address this challenge, we propose to first construct a \emph{priority strategy} that guides the LLM to select events and assertions from a smaller set with higher priority, which is more relevant to the target functionality. If this selection fails, we then guide the LLM to select events and assertions from a larger set with lower priority. This priority strategy restricts the selection set, enhancing the accuracy of selecting events and assertions. Furthermore, considering potential inaccuracies using LLMs, we propose to validate the events and assertions selected by the LLM, thereby improving the effectiveness of the generated test cases to test the target functionality.

To implement the preceding idea, Concretizer involves three key modules as depicted in \figref{fig:workflow}: the \emph{Event/Assertion Matching}, \emph{Event/Assertion Completion}, and \emph{Test Validation} module. 
Specifically, \toolNameSmall{} utilizes one of existing migration approaches~\cite{zhang2024learning, lin2019test, behrang2019test} to obtain a set of events and assertions derived from the source test cases. These events and assertions, referred to as \emph{privileged events and assertions}, are relevant to test the functionality of the target app.
Based on our designed priority strategy, \toolNameSmall{} initially matches each test step in the general test logic with these selected events and assertions using the \textbf{Event/Assertion Matching} module (see \secref{sec:event/assertion-matching}). If the matching fails, \toolNameSmall{} dynamically explores the target app to identify appropriate events and assertions using the \textbf{Event/Assertion Completion} module (see \secref{sec:event/assertion-selection}). Additionally, the \textbf{Test Validation} module (see \secref{sec:test-validation}) plays a crucial role in validating and correcting any errors that arise. These three modules collaboratively work to select appropriate events and assertions from the target app, ensuring the generated test cases are accurate and effective.

Note that, since existing migration approaches may generate incorrect events and assertions or omit essential ones~\cite{zhao2020fruiter} when migrating GUI test cases, we only treat the events and assertions generated by existing migration approaches as privileged but not as the ground-truth.

\subsubsection{Event/Assertion Matching}
\label{sec:event/assertion-matching}

This module aims to match the privileged events and assertions with test steps in the general test logic.
By first selecting events and assertions from the small privileged set, rather than from the entire target app, we enhance the LLM to accurately select appropriate events and assertions.
By selecting events and assertions based on each structured and concise test step, rather than a block of texts for the whole functionality, we aid the LLM in systematically deconstructing the target functionality and incrementally generating test cases.

\begin{figure}[t]
	\center
 \includegraphics[scale = 0.4]{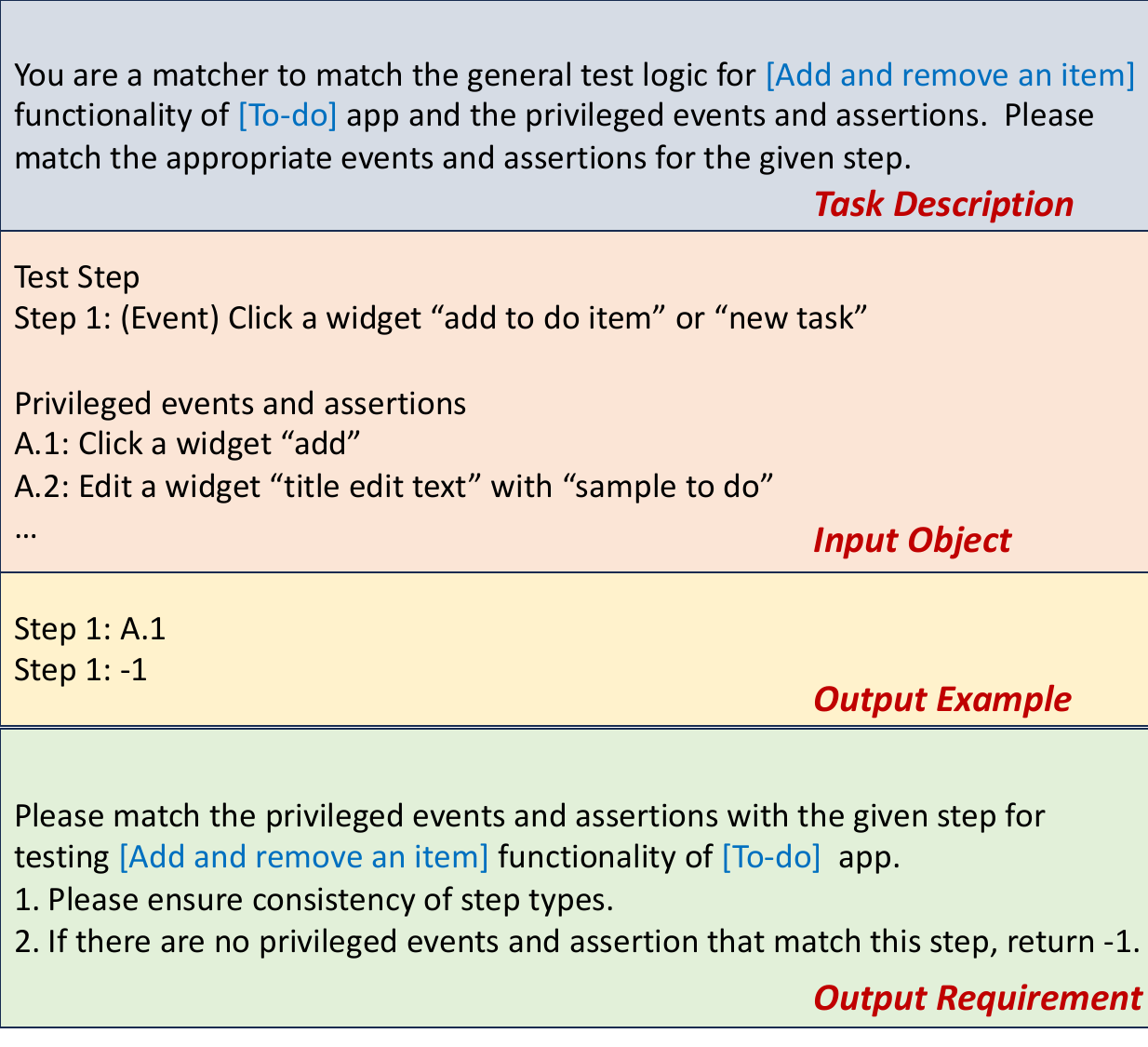}
	\caption{A prompt example for Event/Assertion Matching}
 % \vspace{-5mm}
	\label{fig:prompt-match}
\end{figure}

Specifically, we utilize the LLM to select
events and assertions for each test step in the general test logic. 
The prompt structure used in this module also follows four parts, as shown in \tabref{tab:prompt-template}.
\figref{fig:prompt-match} is an example for this module.

%figure

The main differences between the prompt design in the TL Summarization module and this module are the ``input object'' and ``output requirement''.
First, \emph{input object} displays the test step to match and the privileged events and assertions that have not been matched by the preceding test steps. Second, \emph{output requirement} emphasizes two requirements.
Specifically, the first requirement emphasizes that not every test step needs to match an event or assertion of the privileged events and assertions because these privileged events and assertions may not fully cover the target functionality. Thus, the LLM should return an unmatched indicator (e.g., ``-1'' in our design) if no corresponding event or assertion is found.
The second requirement emphasizes the need for type alignment. For example, a test step with the event type should only match events but not assertions, thus avoiding incorrect matching by the LLM.

\subsubsection{Event/Assertion Completion}
\label{sec:event/assertion-selection}
This module aims to select events and assertions in the target app when the privileged events and assertions cannot be matched to a given test step.
The key parts are \emph{state description}, \emph{event selection}, and \emph{assertion generation}.

%figure
\begin{figure}[t]
	\center
 \includegraphics[scale = 0.4]{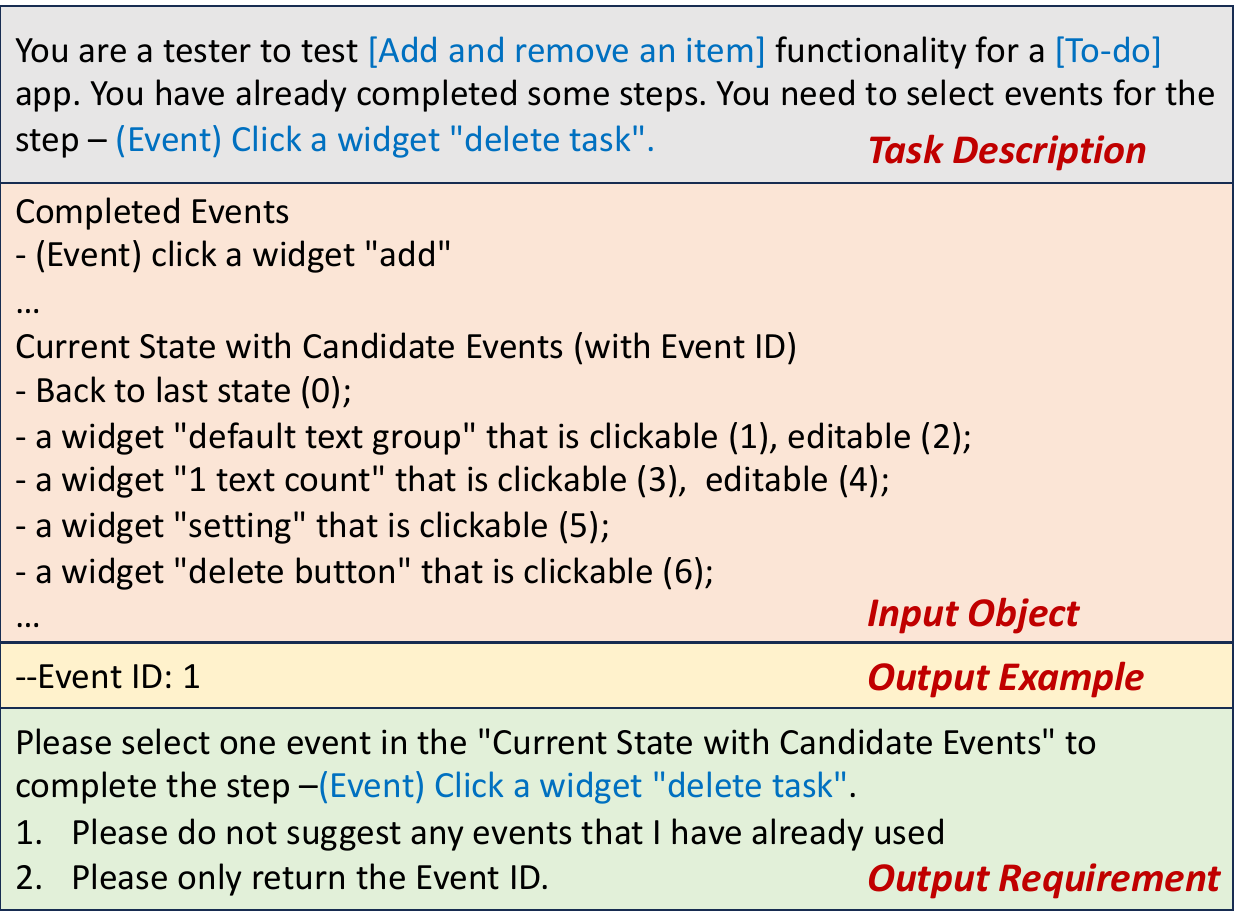}
	\caption{A prompt example for Event/Assertion Completion}
	\label{fig:prompt-generator}
 % \vspace{-2mm}
\end{figure}

\underline{State description.}
To select events and assertions, the LLM needs to understand the semantics of GUI widgets and the actions that these widgets are capable of implementing in the target app. 
\toolNameSmall{} converts the current GUI state of the target app into a natural language description to aid the LLM's understanding. For each widget, \toolNameSmall{} lists the widget semantics and the associated actions. 
The widget semantics are represented by three key attributes, i.e., \emph{text}, \emph{content-desc}, and \emph{resource-id}.
For all widgets within a given state, \toolNameSmall{} organizes them according to their spatial locations, adhering to a trajectory from the top-left to the bottom-right of the state. This sorting maintains a natural and intuitive flow in the state description. 
For instance, the ``input object'' of \figref{fig:prompt-generator} provides the state description for the ``S4'' state in \figref{fig:motivation-example}.

\underline{Event selection.}
Given a test step with event type (referred to as an event step) from the general test logic and the current state, \toolNameSmall{} selects the appropriate events in the target app to complete this step with the LLM.
\figref{fig:prompt-generator} is an example prompt for this part.  
The ``Input Object'' of this prompt not only provides a description of the current state but also displays the selected events from the preceding steps to avoid duplicated selection.

Specifically, \toolNameSmall{} first generates a prompt (e.g., \figref{fig:prompt-generator}) including an event step and a state description of the current state to the LLM.
The LLM returns an event ID from the state description. 
\toolNameSmall{} then executes the event corresponding to that event ID for updating the current GUI state and performs a test validation, as detailed in the Test Validation module (see \secref{sec:test-validation}).
When the test validation passes, \toolNameSmall{} incorporates the event into the GUI test case.
Note that, if the LLM cannot select the appropriate events for the current test step after trying $Max\_selection$ events (i.e., the test validation cannot pass), \toolNameSmall{} skips this test step.
This discrepancy may arise because the extracted general test logic is designed to be broadly comprehensive for the functionality, aiming to cover various possible implementations. However, the specific implementation of the target app might not incorporate every step outlined in this general test logic.

\underline{Assertion generation.} 
An assertion includes a widget and a condition.
GUI testing primarily involves two types of conditions~\cite{behrang2019test, lin2019test}. The first type involves checking the presence of a widget in the current state (e.g., A1 in \figref{fig:motivation-example}), while the second type involves checking the disappearance of a widget in the current state that appears in a preceding state (e.g., A2 in \figref{fig:motivation-example}). As the widget for the second type of assertions does not appear in the current state, we cannot select assertions in the same way as event selection.

\begin{figure}[t]
	\center
 \includegraphics[scale = 0.4]{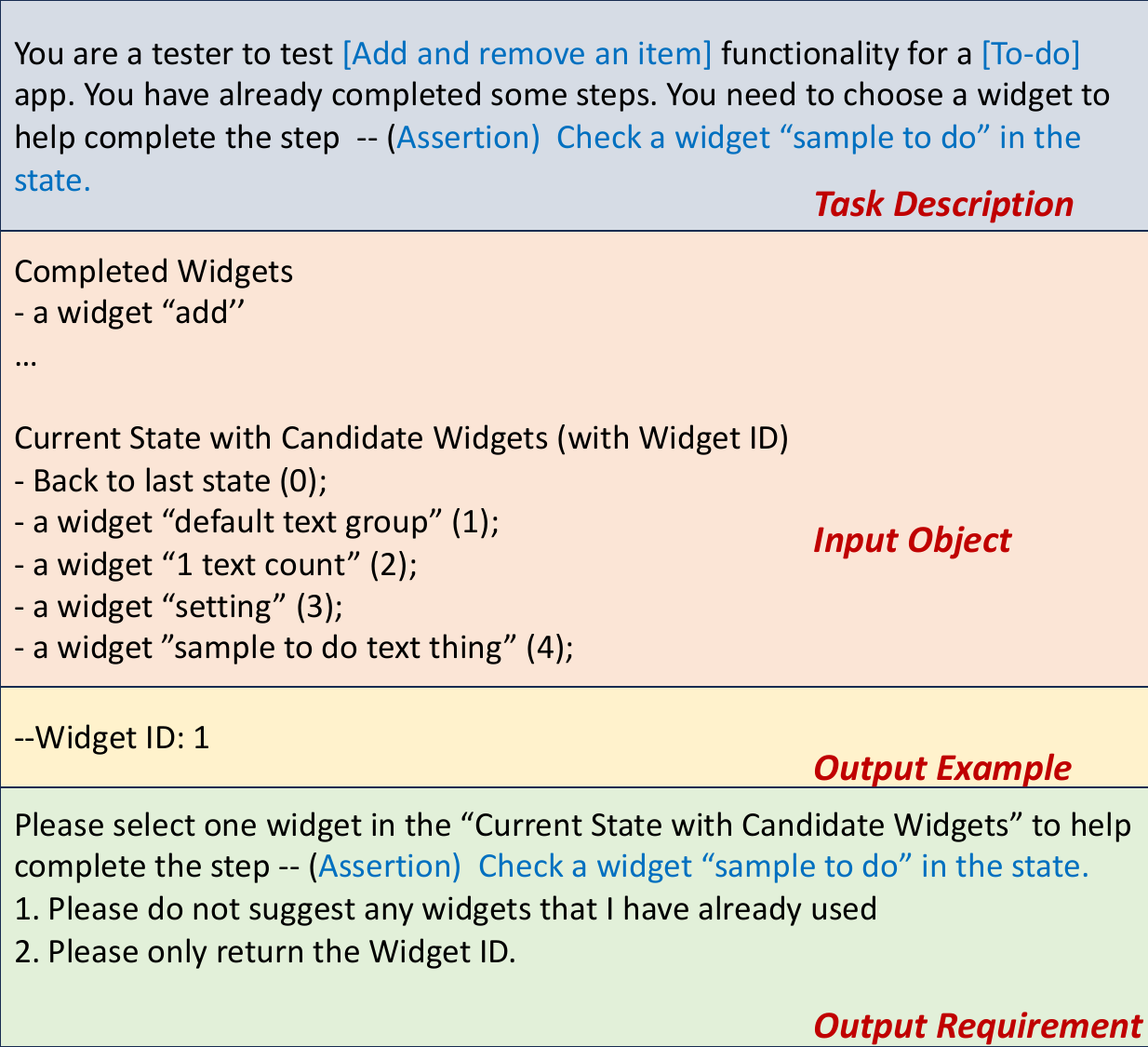}
	\caption{A prompt example for widget selection}
 % \vspace{-5mm}
	\label{fig:prompt-widget}
\end{figure}

To address this problem, given a test step with the assertion type (referred to as an \emph{assertion step}) from the general test logic and the current state, \toolNameSmall{} first selects the appropriate widget in the target app and utilizes the widget to generate the corresponding assertion. 
Unlike event selection, this prompt removes the widget-associated actions, allowing the LLM to focus solely on the widget semantics.
A related prompt is \figref{fig:prompt-widget}.

%figure

Specifically, for the first type of condition, \toolNameSmall{} inputs the current state and the current test step into the LLM. The LLM then selects the appropriate widget based on the step description.
For the second type of condition, since the widget is not present in the current state but appears in a previous state during the generation of this GUI test case, \toolNameSmall{} backtracks from the current state to the previous state to identify the widget.
After \toolNameSmall{} identifies the appropriate widget, it generates a corresponding assertion based on the widget and the condition.
If the LLM cannot select the appropriate widgets for the current step after trying $Max\_selection$ widgets, \toolNameSmall{} skips this step.

% \vspace{-0.2cm}
\subsubsection{Test Validation}
\label{sec:test-validation}
Both the Event/Assertion Matching module and Event/Assertion Completion module utilize LLMs, making their outputs susceptible to the inaccuracies inherent in LLMs.
To address these issues, we design four rules to identify common issues in these modules and provide feedback to LLMs for repairing them. The overview introduction is shown in \tabref{tab:validation-rule}.

\textbf{Validation on the Event/Assertion Matching module.} 
We check the type and the content of the outputs in this module.
There are two common issues (i.e., incorrect type and irrelevant matching) in this module.

\underline{Incorrect type.} The LLM may incorrectly match assertions with an event step, or vice versa, leading to type faults. To identify this issue, \toolNameSmall{} compares the type of each test step with the corresponding matched events/assertions. Upon identifying this issue, \toolNameSmall{} generates a feedback prompt to the Matching module for repairing it. The feedback prompt is ``\emph{The type of [step] and the corresponding events/assertions are not aligned. Please re-match this step}''.

\underline{Irrelevant matching.} 
The outputs of the Matching module may include the events and assertions that do not appear in the privileged set. 
To identify this issue, \toolNameSmall{} compares the outputs of the Matching module with the privileged events and assertions.
Upon identifying this issue, \toolNameSmall{} sends the corresponding feedback to the Matching module: ``\emph{The [Step] matches new events and assertions. Please re-match this step}''.
 
\textbf{Validation on the Event/Assertion Completion module.} 
Below are two common issues in this module.

\underline{Completion checking.} One challenge in using LLMs for GUI test migration is the difficulty in determining the completion of specific test steps due to the comprehension biases~\cite{wen2023empowering}. The LLM often continues to select events and assertions for a test step without termination, resulting in the generation of irrelevant events that deviate from the target functionality. 
To address this issue, we design a checking mechanism to validate whether the current test step has been completed after each event or assertion selection. 
The checking prompt is  ``\emph{Based on [Events] or [Assertions] you generated for [Step], I would like to confirm if [Step] has been successfully completed. Please provide a response in just yes or no}''.
If this mechanism passes (i.e., a response of ``yes''), \toolNameSmall{} proceeds to select events and assertions for the next test step.

\underline{Incorrect format.} The outputs generated by the Completion module do not adhere to the required formats, which may influence \toolNameSmall{} to accurately locate the selected events and assertions. To identify this issue, we compare the outputs of the Completion module with the format of the ``output example''. The feedback is: ``\emph{The [Step] does not adhere to the required formats. Please re-generate this step with the provided format}''.

\section{Evaluation}\label{sec:evaluation}
To evaluate the effectiveness of \toolNameSmall{}, we aim to answer the following research questions:

\textbf{RQ1:} How effective is \toolNameSmall{} compared with the baselines?

\textbf{RQ2:} How do \toolNameSmall{}'s main techniques affect the GUI test migration?

\textbf{RQ3:} How efficient is \toolNameSmall{} compared with the baselines?

\textbf{RQ4:} How useful is \toolNameSmall{} in new apps?

\subsection{Experimental Setup}
\label{sec:experimental-setup}

\textbf{Experimental objects.} 
We select two widely-used datasets for evaluating GUI test migration, which are the FrUITeR dataset~\cite{fruiter-dataset} and the Lin dataset~\cite{craftdroid-dataset}. These datasets include a variety of complex industrial apps (e.g., ABC News~\cite{abc-news} and Firefox Browser~\cite{firefox}), which may help to evaluate \toolNameSmall{} in real-world scenarios. Both these two datasets provide apps along with test cases for target functionalities written by developers (i.e., the ground-truth test cases). The test cases in the Lin dataset contain both events and assertions, while those in the FrUITeR dataset contain only events. We consider all the installable apps and executable test cases provided by the two datasets as our experimental objects.
For the entire evaluation experiments, we evaluate \toolNameSmall{} and related approaches using 31 apps, 34 functionalities, and 123 test cases. \tabref{tab:benchmark} presents basic statistics of our experimental objects. 
Notably, different apps within the same category may share the same functionalities. For example, the five apps in the Browser category all share two same functionalities, which require 10 corresponding test cases. 

Test case migration involves migrating source test cases from source apps to target apps within the same app category. The source apps and target apps are distinct. 
In the experimental setup of \toolNameSmall{}, we employ an iterative methodology in which one app from a selected dataset is designated as the target app, while the remaining apps within the same category of the same dataset are utilized as source apps for migration.
For example, there are five apps in the News category of the FrUITeR dataset (see \tabref{tab:benchmark}). In our experiments, we iterate over each of the five apps, designating one as the target app, while the test cases from the other four apps are used as the source test cases and these four apps are used as the source apps.  This iterative methodology enables a comprehensive evaluation

\begin{table}[t]
  \centering
  \caption{Statistics of experimental objects}
    \begin{tabular}{c|l|r|r|r|r|r|r}
    \toprule
    \textbf{Dataset} & \multicolumn{1}{c|}{\textbf{Category}} & \multicolumn{1}{c|}{\textbf{App}} & \multicolumn{1}{c|}{\textbf{Functionality}} & \multicolumn{1}{c|}{\textbf{Test}} & \multicolumn{1}{c|}{\textbf{Event}} & \multicolumn{1}{c|}{\textbf{Assertion}} & \multicolumn{1}{c}{\textbf{Ave\_Size}} \\
    \midrule
    \multirow{3}[4]{*}{FrUITeR} & News  & 5     & 12    & 42    & 112   & -     & 22M \\
          & Shopping & 5     & 12    & 39    & 207   & -     & 20M \\
\cmidrule{2-8}          & Total & 10    & 24    & 81    & 319   & -     & 21M \\
    \midrule
    \multirow{6}[4]{*}{Lin} & Browser & 5     & 2     & 10    & 32    & 20    & 4M \\
          & To-Do & 5     & 2     & 10    & 39    & 15    & 2M \\
          & Shopping & 4     & 2     & 8     & 49    & 26    & 25M \\
          & Mail  & 2     & 2     & 4     & 14    & 12    & 6M \\
          & Calculator & 5     & 2     & 10    & 33    & 10    & 2M \\
\cmidrule{2-8}          & Total & 21    & 10    & 42    & 167   & 83    & 7M \\
    \bottomrule
    \end{tabular}%
  \label{tab:benchmark}%
\end{table}%

\textbf{Baseline approaches.}
There are two categories of approaches that can generate GUI test cases, i.e., migration approaches~\cite{hu2018appflow,lin2019test, zhang2024learning, behrang2019test, liu2023test} and generation approaches~\cite{wen2023droidbot,wen2023empowering, taeb2023axnav, gao2023assistgui}. 
Existing migration approaches migrate source test cases to target apps based on widget mapping.
Generation approaches require a manually crafted test logic as the input, and use LLMs to select appropriate events in the target app based on test logic. 
To comprehensively evaluate \toolNameSmall{}, we employ representative approaches from both categories. Specifically,
we compare \toolNameSmall{} with  TEMdroid~\cite{zhang2024learning} (the state-of-the-art migration approach) and AutoDroid~\cite{wen2023empowering} (the state-of-the-art generation approach).

\begin{figure}[t]
  % \vspace{-1mm}
	\center
 \includegraphics[scale = 0.4]{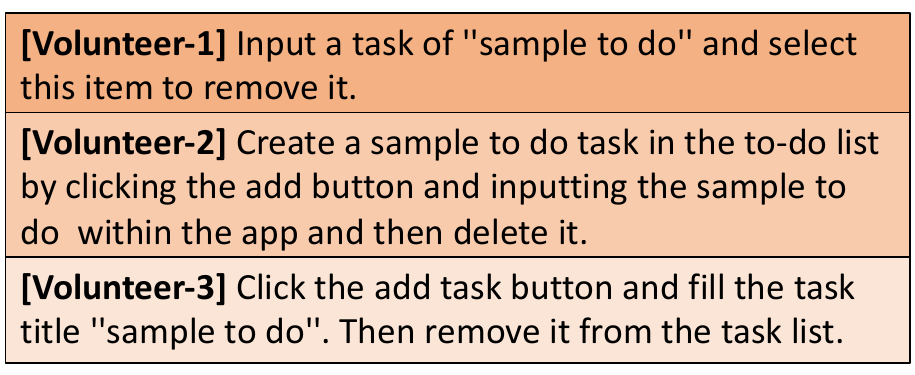}
	\caption{An illustration of AutoDroid inputs}
	\label{fig:autodroid-input}
  % \vspace{-2mm}
\end{figure}

Note that, generation approaches (including AutoDroid) require manually crafted test logics as inputs, but the FrUITeR and the Lin datasets do not provide this information. To compare \toolNameSmall{} with AutoDroid on these two datasets, we invite volunteers to write the necessary test logics. We mitigate the potential influence of different writing styles on the effectiveness of AutoDroid by engaging three volunteers\footnote{None of the volunteers are co-authors of this paper.} with industrial experience in Android programming ranging from 3 to 5 years. For each volunteer, we provide the example descriptions from AutoDroid, the target apps, and the ground-truth test cases for the target functionalities. Each volunteer independently writes the descriptions for all the functionalities to be tested in the two datasets. Since AutoDroid cannot generate assertions, we instruct the volunteers to omit descriptions related to assertions.
For example, AutoDroid utilizes three manually crafted test logics (see \figref{fig:autodroid-input}) as the test logics of \figref{fig:motivation-example}.

\textbf{Evaluation metrics.}
To evaluate \toolNameSmall{} and the baselines, we design three metrics: executable-rate, success-rate, and perfect-rate.

\underline{Executable-rate.} This metric is calculated as the ratio of migrated test cases that can be fully executed ($Test_{exe}$) to the total number of migrated test cases for the target functionalities ($Test_t$).

\begin{equation}
    \textit{Executable-rate} = \textit{Test}_\textit{exe} \,/\, \textit{Test}_\textit{t}
\end{equation}

\underline{Perfect-rate.} This metric is calculated as the ratio of migrated test cases aligning with the ground-truth test cases ($Test_{gt}$) to the total number of migrated test cases for target functionalities ($Test_t$).

\begin{equation}
\textit{Perfect-rate} = \textit{Test}_\textit{gt}\,/\,\textit{Test}_\textit{t}
\end{equation}

\underline{Success-rate.} This metric is calculated as the ratio of migrated test cases that successfully test the target functionalities ($Test_{suc}$) to the total number of migrated test cases for the target functionalities ($Test_t$). Note that, migrated test cases being fully executable is a prerequisite for successful testing the target functionalities. Additionally, 
migrated test cases that align with the ground-truth target test cases are a subset of test cases that successfully test the target functionalities~\cite{zhao2020fruiter,lin2022gui, liu2023test}. This adheres to the purpose of test migration, which seeks to utilize migrated test cases as replacements for manually written target test cases.

\begin{equation}
\textit{Success-rate} = \textit{Test}_\textit{suc}\,/\,\textit{Test}_\textit{t}
\end{equation}

Test cases that successfully test one functionality are not necessarily unique. Consequently, for those test cases that are fully executable but do not align with the ground-truth, we adopt a manual check to further investigate whether these test cases still successfully test their target functionalities. These test cases may include not only all the events and assertions of the ground-truth test cases but also additional events and assertions that do not hinder the testing of the specific functionalities. Specifically, we invite the same three volunteers who have written the test logics for the two datasets (see ``Baseline approaches'' of \secref{sec:experimental-setup}) to help check these additional events and assertions. For each manual check, we provide the volunteers with the generated test case, the additional events or assertions to be checked, the target app, and the corresponding ground-truth test case. Each volunteer independently checks the events and assertions. In cases of disagreement, the volunteers discuss until they reach a consensus.

Note that, the goal of test case migration is to generate test cases that successfully test a functionality of a target app. However, as it is often not feasible to obtain all test cases that successfully test a specific functionality, existing migration approaches~\cite{behrang2019test, zhao2020fruiter, hu2018appflow} typically rely on pre-existing test cases from the datasets that are designed to test the target functionality, referring to them as ``\emph{ground-truth test cases}''.
The ``ground-truth test cases'' serve as a \emph{reference set}, representing a subset of test cases that successfully test the target functionalities. Since it is impossible to exhaustively capture all the ground-truth for the target functionality of a given app, we adopt a combined method that incorporates both automated and manual evaluation.

\textbf{Parameter selection.}
\toolNameSmall{} needs three parameters, which are $Max\_ratio$ used in the Abstractor component, $Max\_selection$ used in the Concretizer component, and $Tem$ used for LLMs.  The determinations of specific parameter values are described as follows.

First, $Max\_ratio$ measures whether the generated test logic remains within a reasonable length. If the generated test logic is too long, it may result in redundancy and irrelevant steps.  
To determine the value of $Max\_ratio$, we compare the length of the general test logic to that of the longest source test case. 
Specifically, we set the minimum value of $Max\_ratio$ to 1, as general test logic typically encompasses or exceeds the scope of the individual source test cases. Additionally, we set the maximum value of $Max\_ratio$ to 2, as the individual test logics of source test cases targeting the same functionality exhibit substantial overlap, with only minor variations between them.
Based on these considerations, we select $Max\_ratio$ values of 1, 1.5, and 2 as reasonable ranges.

Second, $Max\_selection$ defines the maximum number of attempts that the LLM makes to select candidate events for a given test step. The design of this parameter aims to balance testing effectiveness with cost, response time, and the inherent uncertainty of LLM outputs. We set the minimum number of $Max\_selection$ to 1, representing the simplest scenario where the LLM tries only once for each test step. It minimizes cost and response time but may result in suboptimal selection. We set the maximum number of $Max\_selection$ to 3 to control costs and response time. Allowing more than three attempts would significantly increase the cost of LLM invocations and lead to longer response times, which could negatively impact user experience in commercial applications. By selecting 1, 2, and 3 as the range for $Max\_selection$, we achieve a balanced trade-off between cost, efficiency, and the quality of the generated results.

Third, $Tem$ is used to determine the appropriate temperature parameter for the LLM to generate higher-quality test cases. In our experiments, we select the GPT series models~\cite{chatgpt, gpt-4}, which have an official temperature parameter ranging from 0 to 2. At temperature 0, the LLM's output is completely deterministic, which leads to low variability. At temperature 2, the randomness of the outputs is maximized, leading to highly unpredictable results. We aim to generate test cases that are both stable and flexible. Therefore, the extremes of 0 and 2 do not meet our requirements. We select 0.4, 0.8, 1.2, and 1.6 as the candidate values for the temperature parameter.

In summary, the candidate parameters for $Max\_ratio$, $Max\_selection$, and $Tem$ are \{1, 1.5, 2\}, \{1, 2, 3\}, and \{0.4, 0.8, 1.2, 1.6\}, respectively. We randomly select 10\% of the total apps in the Lin dataset as a validation set for parameter selection. After conducting experiments with the validation set, we observe that setting $Max\_ratio$ to 1.5, $Max\_selection$ to 3, and $Tem$ to 0.4 yields the best effectiveness. 
Thus, all the experiments utilize this configuration.

\textbf{Common setting.} 
We implement \toolNameSmall{} in Python to support Android~\cite{android} apps. The experiments are based on a Pixel 3 Emulator running Android 6.0. Some apps require installation in this setup, but MACdroid can adapt to others. For the LLM that \toolNameSmall{} utilizes in evaluation, we select two widely-used models: GPT-3.5~\cite{chatgpt} (i.e., the ``gpt-3.5-turbo-0613'' model used in this evaluation) and GPT-4.0~\cite{gpt-4} (i.e., the ``gpt-4-0613'' model used in this evaluation) to compare the effectiveness of utilizing different LLMs.

To assess the effectiveness of \toolNameSmall{}, we conduct evaluations across different apps, baselines, and LLMs. Specifically, we evaluate \toolNameSmall{}, TEMdroid~\cite{zhang2024learning}, and AutoDroid~\cite{wen2023empowering}  on the FrUITeR dataset and the Lin dataset, respectively. Both \toolNameSmall{} and AutoDroid need to interact with an LLM, for which we evaluate the effectiveness of these approaches using two different LLMs, i.e., GPT-3.5~\cite{chatgpt} and GPT-4.0~\cite{gpt-4}. 
Due to budget constraints, we evaluate the effectiveness of \toolNameSmall{} and AutoDroid on the two full datasets using GPT-3.5. When using GPT-4.0, we randomly select half apps in each app category of the two datasets. Considering the inherent randomness of LLMs, we run each experiment three times and report the average results.

% \vspace{-0.1cm}
\subsection{RQ1: Effectiveness}
\label{sec:rq1}
We evaluate the effectiveness of \toolNameSmall{}, and compare it with two baselines (i.e., TEMdroid~\cite{zhang2024learning}, and AutoDroid~\cite{wen2023empowering}) using the Executable-rate, Perfect-rate, and Success-rate on the FrUITeR dataset and the Lin dataset, respectively. Additionally, we calculate the statistical significance using the Mann-Whitney U-Test~\cite{mcknight2010mann} and effect size using the Cohen's d~\cite{becker2000effect} between \toolNameSmall{} and the baseline approaches.

\begin{table}[t]
  \centering
  \caption{Overall effectiveness of \toolNameSmall{} and the baselines}
    \begin{tabular}{c|l|l|l|r|r|r}
    \toprule
    \textbf{Dataset} & \multicolumn{1}{c|}{\textbf{Approach}} & \multicolumn{1}{c|}{\textbf{LLM}} & \multicolumn{1}{c|}{\textbf{Vol.}} & \multicolumn{1}{c|}{\textbf{Exec.}} & \multicolumn{1}{c|}{\textbf{Perf.}} & \multicolumn{1}{c}{\textbf{Suc.}} \\
    \midrule
    \multirow{9}[10]{*}{\textbf{FrUITeR}} & \multirow{2}[4]{*}{\textbf{\toolNameSmall{}}} & GPT-3.5 & \multicolumn{1}{r|}{-} & \textbf{100\%} & \textbf{18\%} & \textbf{64\%} \\
\cmidrule{3-7}          &       & GPT-4.0 & \multicolumn{1}{r|}{-} & 100\% & 16\%  & 57\% \\
\cmidrule{2-7}          & \multirow{6}[4]{*}{\textbf{AutoDroid}} & \multirow{3}[2]{*}{GPT-3.5} & V-1   & 100\% & 1\%   & 7\% \\
          &       &       & V-2   & 100\% & 2\%   & 9\% \\
          &       &       & V-3   & 100\% & 1\%   & 6\% \\
\cmidrule{3-7}          &       & \multirow{3}[2]{*}{GPT-4.0} & V-1   & 100\% & 3\%   & 11\% \\
          &       &       & V-2   & 100\% & 5\%   & 13\% \\
          &       &       & V-3   & 100\% & 5\%   & 12\% \\
\cmidrule{2-7}          & \textbf{TEMdroid} & \multicolumn{1}{r|}{-} & \multicolumn{1}{r|}{-} & 64\%  & 14\%  & 22\% \\
    \midrule
    \multirow{11}[14]{*}{\textbf{Lin}} & \multirow{2}[4]{*}{\textbf{\toolNameSmall{}}} & GPT-3.5 & \multicolumn{1}{r|}{-} & \textbf{100\%} & \textbf{57\%} & \textbf{75\%} \\
\cmidrule{3-7}          &       & GPT-4.0 & \multicolumn{1}{r|}{-} & 100\% & 68\%  & 77\% \\
\cmidrule{2-7}          & \textbf{TEMdroid} & \multicolumn{1}{r|}{-} & \multicolumn{1}{r|}{-} & 77\%  & 46\%  & 53\% \\
\cmidrule{2-7}          & \multirow{2}[4]{*}{\toolName{*}} & GPT-3.5 & \multicolumn{1}{r|}{-} & \textbf{100\%} & \textbf{61\%} & \textbf{86\%} \\
\cmidrule{3-7}          &       & GPT-4.0 & \multicolumn{1}{r|}{-} & 100\% & 68\%  & 89\% \\
\cmidrule{2-7}          & \multirow{6}[4]{*}{\textbf{AutoDroid}} & \multirow{3}[2]{*}{GPT-3.5} & V-1   & 100\% & 2\%   & 16\% \\
          &       &       & V-2   & 100\% & 3\%   & 13\% \\
          &       &       & V-3   & 100\% & 2\%   & 18\% \\
\cmidrule{3-7}          &       & \multirow{3}[2]{*}{GPT-4.0} & V-1   & 100\% & 8\%   & 18\% \\
          &       &       & V-2   & 100\% & 14\%  & 27\% \\
          &       &       & V-3   & 100\% & 14\%  & 39\% \\
    \bottomrule
    \end{tabular}%
  \label{tab:evaluation-FTDroid}%
% \vspace{-5mm}
\end{table}

% Table generated by Excel2LaTeX from sheet 'significant-summary'
\begin{table}[htbp]
  \centering
  \caption{Effectiveness of \toolNameSmall{} and the baselines by category}
    \begin{tabular}{c|l|r|r|r|r|r|r|r|r|r}
    \toprule
    \multirow{2}[4]{*}{\textbf{Dataset}} & \multicolumn{1}{c|}{\multirow{2}[4]{*}{\textbf{Category}}} & \multicolumn{3}{c|}{\textbf{MACdroid}} & \multicolumn{3}{c|}{\textbf{TEMdroid}} & \multicolumn{3}{c}{\textbf{AutoDroid}} \\
\cmidrule{3-11}          &       & \multicolumn{1}{c|}{\textbf{Exec.}} & \multicolumn{1}{c|}{\textbf{Perf.}} & \multicolumn{1}{c|}{\textbf{Suc.}} & \multicolumn{1}{c|}{\textbf{Exec.}} & \multicolumn{1}{c|}{\textbf{Perf.}} & \multicolumn{1}{c|}{\textbf{Suc.}} & \multicolumn{1}{c|}{\textbf{Exec.}} & \multicolumn{1}{c|}{\textbf{Perf.}} & \multicolumn{1}{c}{\textbf{Suc.}} \\
    \midrule
    \multirow{2}[2]{*}{\textbf{FrUITeR}} & News  & 100\% & 21\%  & 71\%  & 42\%  & 18\%  & 19\%  & 100\% & 1\%   & 6\% \\
          & Shopping & 100\% & 15\%  & 58\%  & 79\%  & 12\%  & 23\%  & 100\% & 2\%   & 8\% \\
    \midrule
    \multirow{5}[2]{*}{\textbf{Lin}} & Browser & 100\% & 55\%  & 100\% & 85\%  & 85\%  & 85\%  & 100\% & 0\%   & 22\% \\
          & To-Do & 100\% & 50\%  & 70\%  & 58\%  & 28\%  & 28\%  & 100\% & 1\%   & 9\% \\
          & Shopping & 100\% & 13\%  & 25\%  & 54\%  & 4\%   & 17\%  & 100\% & 0\%   & 0\% \\
          & Mail  & 100\% & 88\%  & 88\%  & 100\% & 100\% & 100\% & 100\% & 0\%   & 8\% \\
          & Calculator & 100\% & 90\%  & 90\%  & 100\% & 45\%  & 65\%  & 100\% & 9\%   & 32\% \\
    \bottomrule
    \end{tabular}%
  \label{tab:effectiveness-by-category}%
\end{table}%

% Table generated by Excel2LaTeX from sheet 'significant-summary'
\begin{table}[htbp]
  \centering
  \caption{Significant difference and Effect size of MACdroid and the baselines}
    \begin{tabular}{c|l|r|r|r|r}
    \toprule
    \multirow{2}[4]{*}{\textbf{Dataset}} & \multicolumn{1}{c|}{\multirow{2}[4]{*}{\textbf{Category}}} & \multicolumn{2}{c|}{\textbf{MACdroid vs TEMdroid}} & \multicolumn{2}{c}{\textbf{MACdroid vs AutoDroid}} \\
\cmidrule{3-6}          &       & \multicolumn{1}{l|}{\textbf{Significance}} & \multicolumn{1}{l|}{\textbf{Effect size}} & \multicolumn{1}{l|}{\textbf{Significance}} & \multicolumn{1}{l}{\textbf{Effect size}} \\
    \midrule
    \multirow{2}[2]{*}{\textbf{FrUITeR}} & News  & \textcolor[rgb]{ 1,  0,  0}{1.97E-19} & \textcolor[rgb]{ 1,  0,  0}{0.49} & \textcolor[rgb]{ 1,  0,  0}{2.31E-51} & \textcolor[rgb]{ 1,  0,  0}{0.49} \\
          & Shopping & \textcolor[rgb]{ 1,  0,  0}{2.87E-10} & \textcolor[rgb]{ 1,  0,  0}{0.34} & \textcolor[rgb]{ 1,  0,  0}{7.24E-29} & \textcolor[rgb]{ 1,  0,  0}{0.36} \\
    \midrule
    \multirow{5}[2]{*}{\textbf{Lin}} & Browser & \textcolor[rgb]{ 1,  0,  0}{0.03} & 0.13  & \textcolor[rgb]{ 1,  0,  0}{9.44E-14} & \textcolor[rgb]{ 1,  0,  0}{0.58} \\
          & To-Do & \textcolor[rgb]{ 1,  0,  0}{4.65E-03} & \textcolor[rgb]{ 1,  0,  0}{0.36} & \textcolor[rgb]{ 1,  0,  0}{1.59E-11} & \textcolor[rgb]{ 1,  0,  0}{0.46} \\
          & Shopping & 0.49  & 0.07  & \textcolor[rgb]{ 1,  0,  0}{1.36E-05} & 0.18 \\
          & Mails & 0.45  & -0.12 & \textcolor[rgb]{ 1,  0,  0}{5.89E-07} & \textcolor[rgb]{ 1,  0,  0}{0.55} \\
          & Calculator & \textcolor[rgb]{ 1,  0,  0}{0.04} & 0.18  & \textcolor[rgb]{ 1,  0,  0}{2.49E-07} & \textcolor[rgb]{ 1,  0,  0}{0.41} \\
    \bottomrule
    \end{tabular}%
  \label{tab:significant}%
\end{table}%

\textbf{Effectiveness results.} 
\tabref{tab:evaluation-FTDroid} shows the overall effectiveness of the test cases generated by \toolNameSmall{}, TEMdroid, and AutoDroid on the FrUITeR dataset and Lin dataset, using both GPT-3.5 and GPT-4.0 as the LLMs.
For the results related to GPT-3.5, we further analyze the effectiveness of these approaches across different app categories (see \tabref{tab:effectiveness-by-category}) and provide the significance and effect sizes of these approaches by category (see \tabref{tab:significant}).
We count the executable-rate (denoted as ``Exec.''), perfect-rate (denoted as ``Perf.''), and success-rate (denoted as ``Suc.'') of \toolNameSmall{} and baselines. 
We also separately count the impact of the test logics written by the three volunteers (denoted as ``Vol.'') on the effectiveness of AutoDroid.

\underline{Effectiveness on the FrUITeR dataset.} All the test cases generated by \toolNameSmall{} are fully executable (i.e., 100\%). When utilizing GPT-3.5 as the LLM, 64\% of the test cases generated by \toolNameSmall{} successfully test the target functionalities (i.e., success-rate), and 18\% of them align with the corresponding ground-truth (i.e., perfect-rate). These results surpass TEMdroid and AutoDroid by over 191\% in success-rate and 29\% in perfect-rate. 
Note that, when GPT-4.0 is used, the test cases generated by \toolNameSmall{} also show a slight improvement in the success-rate (i.e., 57\% in \tabref{tab:evaluation-FTDroid}) compared to using GPT-3.5, which achieves a 54\% success-rate in the same \emph{half} of the total apps.

\underline{Effectiveness on the Lin dataset.} When using GPT-3.5 as the LLM, 75\% of the test cases generated by \toolNameSmall{} successfully test the target functionalities, and 57\% of them align with the ground-truth test cases. These results outperform TEMdroid by 42\% in success-rate and 24\% in perfect-rate. 
We also separately calculate the accuracy of generated assertions by \toolNameSmall{} and TEMdroid at the case level. MACdroid achieves the assertion accuracy of 83\%, compared to 75\% for TEMdroid.

The test cases in the Lin dataset include both events and assertions, but AutoDroid cannot generate assertions. To fairly compare \toolNameSmall{} with AutoDroid on the Lin dataset, we evaluate the test cases generated by \toolNameSmall{} and AutoDroid without considering the generated assertions. 
In this scenario, \toolNameSmall{} (denoted as ``\toolNameplain{*}'' ) outperforms AutoDroid by more than 378\% in success-rate and 1933\% in perfect-rate. 
Additionally, switching from GPT-3.5 to GPT-4.0 also slightly improves the effectiveness of \toolNameSmall{}.

\underline{Statistical analysis.} 
\tabref{tab:significant} presents the significant differences and effect sizes between MACdroid and TEMdroid, as well as between MACdroid and AutoDroid. In this table, red numbers indicate statistically significant differences or large effect sizes, while black numbers indicate no significant difference or not large effect sizes. As can be seen, MACdroid shows statistically significant differences compared to both TEMdroid and AutoDroid, with large effect sizes.

Note that, all results from the FrUITeR dataset exhibit statistical significance and large effect sizes, whereas not all results from the Lin dataset show similar trends. The reason for this phenomenon lies in the difference in the number of test cases per category in the two datasets. Specifically, the FrUITeR dataset contains an average of 41 different test cases for each category, while the Lin dataset contains only an average of 8 test cases per category. As statistical analysis requires a sufficient number of samples, the small number of test cases per category for the Lin dataset are difficult to show significant differences.

\textbf{Result analysis.}
We have several findings from \tabref{tab:evaluation-FTDroid}.

First, compared to TEMdroid, \toolNameSmall{} significantly improves the success-rate (e.g., 22\% vs. 64\% on the FrUITeR dataset). 
Existing migration approaches only migrate a test case to the target app, but the same functionality can be implemented differently across various apps. As a result, the migrated test case may only partially test the functionality of the target app. In contrast, \toolNameSmall{} extracts the general test logic from multiple test cases to provide a comprehensive perspective of the target functionality, which is conducive to completely testing the target functionality.

Second, the test cases generated by TEMdroid may not be fully executable (see the column of ``Exec.''). This is because existing migration approaches migrate test cases based on widget mapping, but the migrated events and assertions may miss some connection events in the target app, making the generated test case not fully executable. 
In contrast, both \toolNameSmall{} and AutoDroid generate test cases by incrementally selecting and executing events within the target app, thereby ensuring that all connection events are retained and the generated test cases are fully executable.

Third, compared to AutoDroid, \toolNameSmall{} significantly improves the perfect-rate (e.g., 3\% vs. 61\% on the Lin dataset). Existing generation approaches struggle to generate perfect test cases because the provided test logics may be vague and ambiguous, making it challenging for the LLM to determine whether the functionalities have been fully tested, thus including irrelevant events. 
In contrast, \toolNameSmall{} splits a whole test logic into a sequence of structured test steps, guides the LLM to generate test cases step-by-step, and verifies the completion of each step. This approach effectively reduces the generation of irrelevant events and assertions, facilitating the generation of perfect test cases.

Fourth, even when AutoDroid utilizes GPT-4.0, it still does not perform as well as \toolNameSmall{} utilizing GPT-3.5. This situation indicates that simply upgrading the LLM does not substantially enhance the generated test cases. Instead, providing the LLM with high-quality test logics, restricting the input candidates for the LLM, and repairing potential errors are important for improvement.

Fifth, the effectiveness of generated test cases for AutoDroid varies depending on the test logics written by different volunteers. For example, the success-rates of AutoDroid on the FrUITeR dataset are 7\%, 9\%, and 6\%, respectively when utilizing the descriptions from the three volunteers. 
This variability underscores the impact of human factors on the robustness of existing generation approaches. Instead, the general test logic extracted by \toolNameSmall{} not only automates the generation of test logics but also ensures stable quality for the LLM to generate test cases.

Sixth, according to \tabref{tab:effectiveness-by-category} we can  observe that MACdroid achieves high effectiveness across different app categories and outperforms the baselines, demonstrating the robustness of MACdroid.

Seventh, the accuracy of assertion generation is also evaluated at the test case level, as the test case serves as the fundamental unit for functional testing. The accuracy is defined as the condition where all assertions generated for a test case are completely consistent with the assertions in the ground-truth test cases. Assertions play a crucial role in functional testing~\cite{korel1996assertion, foster2004assertion}. Among the test cases generated by MACdroid, 83\% of them include assertions that successfully meet the requirements for testing the corresponding functionalities.
This result indicates that MACdroid is capable of generating high-quality assertions effectively and demonstrates strong effectiveness in test case migration compared to related approaches.

\begin{figure}[t]
  % \vspace{-1mm}
	\center
 \includegraphics[scale = 0.25]{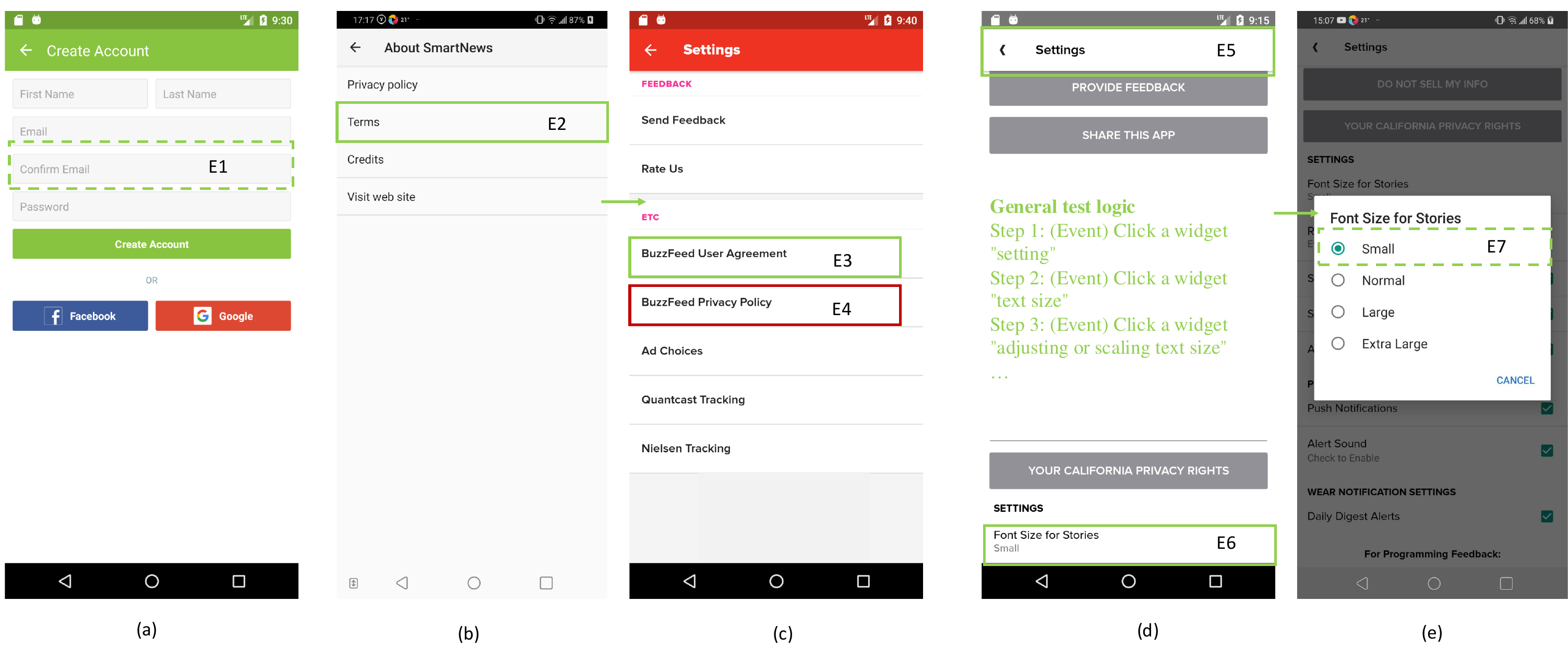}
	\caption{Failure case study}
	\label{fig:failure-case-study}
  % \vspace{-2mm}
\end{figure}

% \fei{better throw the percentages for each portion}
\textbf{Failure Analysis.} 
To understand the weaknesses of \toolNameSmall{}, we manually analyze all failure test cases and identify three main reasons. 
We select three examples (see \figref{fig:failure-case-study}) to introduce the failure reasons of MACdroid and also to provide the potential solutions to address these failures.

First, a limited number of source test cases may not cover all the necessary test steps for the target functionality, leading to the generated test case not fully testing this functionality. 
For example, the state (a) of \figref{fig:failure-case-study} shows a registration functionality. MACdroid does not generate the necessary event (i.e., E1) because the source test cases do not include steps related to confirmation.
Increasing the diversity of source test cases, while leveraging the general knowledge of LLMs to supplement necessary steps for general test logic, may potentially help address this problem.

Second, the Event/Assertion Matching module may incorrectly match events and assertions to the corresponding test steps, resulting in the generated test cases containing incorrect events and assertions. 
For example, states (b) and (c) of \figref{fig:failure-case-study} illustrate a functionality designed to test the terms. The state (b) represents a part of the source test case, while the state (c) shows the corresponding part of the target test case.
Given the E2 event from the source test case, the Event/Assertion Matching module incorrectly matches E4 event instead of the correct event, i.e., E3. This happens because the content descriptions of E4 (i.e., privacy policy) and E2 (i.e., terms) are related, both falling under the broader agreement category, making the matching process prone to error. Incorporating the state information into the matching process and utilizing more accurate matching algorithms may help improve the effectiveness of the Event/Assertion matching module.

Third, the Event/Assertion Completion module may also select incorrect events and assertions or do not select any events and assertions, impacting the generated test cases to successfully test the target functionality.
We illustrate this issue using states (d) and (e) in \figref{fig:failure-case-study}, which depict the functionality of changing text size.
In this case, the Event/Assertion Completion module follows a general test logic (highlighted in green in the state (d)) to generate events E5 and E6 for the first two steps of the general test logic. However, the module determines that no events are suitable for Step 3, leading to the omission of an event selection for this step. As a result, the E7 event is missing from the final generated test case.
Enhancing the Completion module with more self-learning capabilities, and ensuring it fully understands the semantics behind each step of the general test logic, could help mitigate this issue.

\finding{\textbf{Answer to RQ1:} \toolNameSmall{} is effective and substantially outperforms the baselines in GUI test migration.}

% \vspace{-0.2cm}
\subsection{RQ2: Main Techniques}
We evaluate the contributions of the main techniques used in \toolNameSmall{} based on metric evaluation and statistical analysis on the FrUITeR dataset. The LLM used is GPT-3.5.

\textbf{Experimental setting.} There are four experiments.
First, we evaluate the effectiveness of Abstractor in \toolNameSmall{}. Specifically, we modify AutoDroid by replacing its manually crafted test logics with the extracted general test logics by Abstractor, denoted as ``AutoDroid (with Abstractor)''. We then compare the effectiveness of the original AutoDroid with ``AutoDroid (with Abstractor)''.
Second, we investigate the effectiveness of Concretizer in \toolNameSmall{}. 
Specifically, we compare the test cases generated by ``AutoDroid (with Abstractor)'' with those generated by the original \toolNameSmall{}, as they both use the same test logics but different generation approaches.
Third, we investigate the effectiveness of the validation modules used in the Abstractor component and the Concretizer component of \toolNameSmall{}. 
Specifically, we compare the test cases generated by \toolNameSmall{} without these validation modules (denoted as ``\toolNameSmall{} (without Validation)'') with those generated by the original \toolNameSmall{}.
Fourth, we investigate the robustness of \toolNameSmall{} when using different sources of privileged events and assertions (i.e., different migration approaches). 
Specifically, we select AppFlow~\cite{hu2018appflow}, a representative migration approach that has been extensively compared in numerous studies~\cite{zhao2020fruiter, zhang2024learning, mariani2021semantic}.
Then, we compare the test cases generated by the original \toolNameSmall{} that uses privileged events and assertions from TEMdroid~\cite{zhang2024learning} with the results generated by \toolNameSmall{} (denoted as ``\toolNameSmall{} (with AppFlow)'') that uses privileged events and assertions from AppFlow~\cite{hu2018appflow}.

% Table generated by Excel2LaTeX from sheet 'RQ3'
\begin{table}[t]
  \centering
  % \small
  \caption{Effectiveness of main techniques in \toolNameSmall{}}
    \begin{tabular}{l|r|r|r}
    \toprule
    \textbf{Approach} & \multicolumn{1}{l|}{\textbf{Executable}} & \multicolumn{1}{l|}{\textbf{Perfect}} & \multicolumn{1}{l}{\textbf{Success}} \\
    \midrule
    \toolNameSmall{} & \textbf{100\%} & \textbf{18\%}  & \textbf{64\%} \\
    AutoDroid & 100\% & 2\%   & 9\% \\
    \midrule
    AutoDroid (with Abstractor) & 100\% & 9\%   & 27\% \\
    \toolNameSmall{} (without Validation) & 100\% & 4\%   & 31\% \\
    \toolNameSmall{} (with AppFlow) & 100\% & 15\%  & 59\% \\
    \bottomrule
    \end{tabular}%
  \label{tab:ablation-study}%
\end{table}%

% Table generated by Excel2LaTeX from sheet 'significant-summary'
\begin{table}[t]
  \centering
  \caption{Significant difference and Effect size of MACdroid and the main techniques}
    \begin{tabular}{l|r|r}
    \toprule
    \textbf{Approach} & \multicolumn{1}{l|}{\textbf{Significance}} & \multicolumn{1}{l}{\textbf{Effect size}} \\
    \midrule
    AutoDroid (with Abstractor) vs AutoDroid & \textcolor[rgb]{ 1,  0,  0}{5.79E-12} & 0.12  \\
    MACdroid vs AutoDroid (with Abstractor) & \textcolor[rgb]{ 1,  0,  0}{4.67E-21} & \textcolor[rgb]{ 1,  0,  0}{ 0.37} \\
    MACdroid vs MACdroid (without Validation) & \textcolor[rgb]{ 1,  0,  0}{1.21E-07} &  0.29 \\
    MACdroid vs MACdroid (with AppFlow) & 0.38  & 0.04  \\
    \bottomrule
    \end{tabular}%
  \label{tab:significant-rq2}%
\end{table}%

\textbf{Effectiveness results.}
\tabref{tab:ablation-study} presents the executable-rate, perfect-rate, and success-rate of this evaluation.
First, by comparing the second row with the third row, we observe that using the general test logics generated by the Abstractor component significantly improves the effectiveness of AutoDroid (e.g., 27\% vs. 9\% in the success-rate). This result indicates that the general test logics generated by the Abstractor component provide better guidance for the LLM compared to manually crafted test logics, making the LLM easier to understand the functionalities being tested. 
Second, by comparing the first row with the third row, we observe that even with the general test logics, AutoDroid does not perform as well as \toolNameSmall{} (e.g., 27\% vs. 64\% in the success-rate). This result suggests that the priority strategy, along with the test validation mechanisms, are also crucial for generating high-quality test cases.
Third, a comparison between the first and fourth rows reveals that the LLM tends to generate some inaccurate results, highlighting the importance of the validation modules used in the Abstractor component and the Concretizer component (e.g., 64\% vs. 31\% in the success-rate).
Fourth, a comparison between the first and fifth rows indicates that \toolNameSmall{} is reliable and robust when employing various sources of privileged events and assertions (e.g., 64\% vs. 59\% in the success-rate).

\textbf{Statistical analysis.} \tabref{tab:significant-rq2} presents the significant differences and effect sizes for the following four comparisons, which are AutoDroid (with Abstractor) versus the original AutoDroid, the original MACdroid versus AutoDroid (with Abstractor), the original MACdroid versus MACdroid (without Validation), and the original MACdroid versus MACdroid (with AppFlow).

In \tabref{tab:significant-rq2}, the red numbers indicate significant differences or large effect sizes, while the black numbers represent no significant differences or not large effect sizes. The statistical results show that the Abstractor component, Concretizer component, and the validation module are crucial in MACdroid. When these modules are removed, significant differences are observed compared to the original version. All results in the fourth row are black, indicating no significant differences when changing different privileged events and assertions. This result also demonstrates that the MACdroid approach exhibits high reliability and robustness, not relying on specific sources of privileged events and assertions.

\finding{\textbf{Answer to RQ2:} \toolNameSmall{}'s main techniques substantially contribute to GUI test migration.}

\subsection{RQ3: Efficiency}

To assess the efficiency of \toolNameSmall{}, we evaluate the execution time and the token usage of \toolNameSmall{}, TEMdroid~\cite{zhang2024learning}, and AutoDroid~\cite{wen2023empowering} on the FrUITeR dataset and the Lin dataset, respectively.

% Table generated by Excel2LaTeX from sheet 'RQ2'
\begin{table}[t]
  \centering
  \caption{Efficiency of \toolNameSmall{} and the baselines}
% Table generated by Excel2LaTeX from sheet 'cost'
    \begin{tabular}{l|r|r|r|r|r|r}
    \toprule
    \multicolumn{1}{c|}{\multirow{2}[4]{*}{\textbf{Dataset}}} & \multicolumn{2}{c|}{\textbf{MACdroid}} & \multicolumn{2}{c|}{\textbf{TEMdroid}} & \multicolumn{2}{c}{\textbf{AutoDroid}} \\
\cmidrule{2-7}          & \multicolumn{1}{l|}{\textbf{Runtime}} & \multicolumn{1}{l|}{\textbf{Token}} & \multicolumn{1}{l|}{\textbf{Runtime}} & \textbf{Token} & \multicolumn{1}{l|}{\textbf{Runtime}} & \multicolumn{1}{l}{\textbf{Token}} \\
    \midrule
    \textbf{FrUITeR} & 5.1   & 6378  & 9.6   & 18     & 5.5   & 6496 \\
    \textbf{Lin} & 4.7   & 8256  & 8.8   & 14     & 4.5   & 9425 \\
    \bottomrule
    \end{tabular}%
  \label{tab:runtime-information}%
\end{table}%

\textbf{Efficiency results.}
\tabref{tab:runtime-information} shows the average runtime and token-usage per test case for \toolNameSmall{}, TEMdroid, and AutoDroid. The average runtime for \toolNameSmall{} per test case is 5.1 minutes on the FrUITeR dataset and 4.7 minutes on the Lin dataset. These results are faster than TEMdroid and comparable to those of AutoDroid.
The average token usage for \toolNameSmall{} per test case is 6378 on the FrUITeR dataset and 8256 on the Lin dataset, which is smaller than that of AutoDroid. Note that, TEMdroid fine-tunes its own model based on BERT~\cite{bert-implement} instead of utilizing LLMs, resulting in the lowest token usage.

The time cost of \toolNameSmall{} is primarily influenced by its two components, i.e., Abstractor and Concretizer. 
For a target functionality, \toolNameSmall{} spends less than 0.5 minutes on average when generating a general test logic using the Abstractor component. The majority of the time is consumed by the Concretizer component in generating test cases. This is because Abstractor interacts with the LLM fewer times than Concretizer does. Ideally, the Abstractor component only needs to interact with the LLM twice (once to generate the test logic and the other to validate it). However, the Concretizer component requires multiple interactions with the LLM to select events for each test step, including matching, completion, and validation. Reducing the number of interactions between the Concretizer component and the LLM would significantly improve \toolNameSmall{}'s efficiency.

\finding{\textbf{Answer to RQ3:} \toolNameSmall{}'s efficiency is comparable to the baselines.}

\subsection{RQ4: Usefulness}
To further evaluate the usefulness of \toolNameSmall{}, we conduct a study evaluating \toolNameSmall{} in new apps. The LLM used in this evaluation is GPT-3.5.

\textbf{Experimental objects.} 
We leverage a new dataset~\cite{TEMdroid-dataset}, referred to as the TEM dataset in this study, which was developed as part of the research on TEMdroid~\cite{zhang2024learning}. The TEM dataset incorporates all the apps and test cases from the Lin dataset~\cite{craftdroid-dataset} as source apps and source test cases.  
These source test cases include both events and assertions, which enables the evaluation of \toolNameSmall{}'s effectiveness in generating comprehensive test cases.
The TEM dataset includes five new target apps, which are popular apps in the Google Play Store~\cite{google-play}, with each app belonging to a distinct category in the Lin dataset. Note that, the TEM dataset does not provide the ground-truth test cases for the target apps.

\textbf{Evaluation process.} To assess the effectiveness of \toolNameSmall{}, we adopt a similar evaluation process as described in \secref{sec:experimental-setup} and engage the same three volunteers. Since the TEM dataset does not include ground-truth test cases, we are unable to provide the volunteers with the ground-truth target test cases or evaluate the perfect-rate metric. 
To address this issue, we provide the volunteers with ground-truth source test cases for the same functionality, helping them gain a clearer understanding of the target functionality.
Thus, the evaluation metrics for this study are executable-rate and success-rate.

\textbf{Usefulness results.} \tabref{tab:new-app} presents the executable-rate and success-rate of the test cases migrated by \toolNameSmall{} on the TEM dataset. In this study, \toolNameSmall{} achieves an executable-rate of 100\% and a success-rate of 73\%. From \tabref{tab:new-app} we observe that \toolNameSmall{} demonstrates significant effectiveness across different app categories, highlighting its satisfactory usefulness in new apps.

\finding{\textbf{Answer to RQ4:} \toolNameSmall{} shows satisfactory usefulness in new apps.}

% Table generated by Excel2LaTeX from sheet 'new-app'
\begin{table}[t]
  \centering
  \caption{Results of the usefulness study in \toolNameSmall{}}
    \begin{tabular}{l|r|r}
    \toprule
    \textbf{App} & \multicolumn{1}{l|}{\textbf{Executable}} & \multicolumn{1}{l}{\textbf{Success}} \\
    \midrule
    Web Browser & 100\% & 67\% \\
    Done  & 100\% & 83\% \\
    Fivemiles & 100\% & 50\% \\
    Pro Mail & 100\% & 83\% \\
    Tip Calculator & 100\% & 83\% \\
    \midrule
    \textbf{Average} & \textbf{100\%} & \textbf{73\%} \\
    \bottomrule
    \end{tabular}%
  \label{tab:new-app}%
\end{table}%

% \vspace{-0.2cm}
\subsection{Threats to Validity}

A possible threat to external validity is the generalizability to other datasets. To mitigate this threat, we use the largest number of apps and app categories compared with related work. 
Moreover, we use all the popular datasets in test case migration, i.e., the Lin dataset~\cite{lin2019test} and the FrUITeR dataset~\cite{fruiter-dataset}, as evaluation benchmarks. These datasets include a variety of complex industrial apps (e.g., ABC News~\cite{abc-news} and Firefox Browser~\cite{firefox}), which may help to evaluate MACdroid and the baselines in real-world scenarios.

A possible threat to internal validity is the possible mistakes involved in our implementation and experiments.
To mitigate this threat, we manually inspect our results and analyze the test cases that fail to test the target functionalities.
We also publish our implementation and experimental data, which welcome external validation.
As for the human evaluation, we invite three experienced developers and provide them with a clear evaluation process.

A possible threat to construct validity is about evaluation metrics.
To mitigate this threat, we carefully design three metrics aiming at validating the effectiveness of the generated test cases. These metrics offer a reliable and objective basis for assessing the quality of the test cases generated by both the baselines and \toolNameSmall{}.

\section{Discussion}\label{sec:discussion}

\textbf{Collecting and retrieving source test cases.} Migration approaches typically leverage source test cases as input to migrate them into target apps. These approaches reduce the cost of manually writing GUI test cases for the target apps and enable the reuse of existing GUI test cases. \toolNameSmall{} is also a migration approach that significantly outperforms existing approaches (see \secref{sec:rq1}). However, applying migration approaches to industry still needs to consider the collection and retrieval of source test cases.

\underline{Collection.} In real-world scenarios, there are many similar apps and corresponding GUI test cases~\cite{hu2018appflow, mariani2021evolutionary, zhao2020fruiter}. For example, app stores (e.g., Google Play Store~\cite{google-play} and F-Droid~\cite{fdroid}) organize similar apps into the same categories (e.g., shopping and news). Additionally, open-source communities (e.g., GitHub~\cite{github}) host numerous mobile apps with test cases. Leveraging these open-source test cases and existing datasets~\cite{craftdroid-dataset,fruiter-dataset}, collecting apps and GUI test cases becomes a straightforward process.

\underline{Retrieval.} These apps and test cases can be organized by app category and the specific functionalities within each category. Cluster-based algorithms (e.g., K-Means~\cite{krishna1999genetic}) can be used to group test cases based on functionality names and test case annotations. Each cluster contains test cases corresponding to specific functionalities within the same category. Given a target app, its category, and the functionality name to be tested, the closest cluster in the collected dataset can be identified, and all test cases associated with that cluster can be retrieved.

\textbf{Impact and solution of the quality of source test cases.} The quality of source test cases impacts the effectiveness of \toolNameSmall{}. Below we first detail analyze the impact of the high-quality source test cases and the low-quality ones on the effectiveness of \toolNameSmall{}. We further analyze the potential solution to the test case selection.

\underline{Impact.} High-quality source test cases, which are representative of testing a target functionality with minimal app-specific information (e.g., environment-specific details), may enhance \toolNameSmall{}'s effectiveness. Such test cases effectively encapsulate the core logical information required to test a target functionality while minimizing interference from irrelevant app-specific details. As a result, the general test logic derived from high-quality source test cases accurately reflects the essence of the target functionality and can be seamlessly instantiated for the target app.

In contrast, low-quality source test cases, which lack representativeness or include excessive app-specific details, may lead to incomplete or noisy test logic generation. This deficiency may lead to inconsistencies between the generated test logic and the actual requirements for testing the target app. Consequently, the test cases derived from such logic may fail to fully test the target functionality, either by only partially testing it or containing numerous errors, and thus still require manual modification.

\underline{Solution.} There are two ways to potentially improve the quality of source test cases. First, the selection of source test cases could be diversified. By selecting test cases from different scenarios and across various mobile apps that test the same functionality, we may obtain a broader and more accurate representation of how the functionality should be tested. Second, filtering out low-quality test cases is also essential. By filtering out test cases that have incomplete test logics or rely heavily on app-specific information, we can reduce interference from irrelevant details, thereby facilitating the accurate generation of the general test logic and the concrete test case.

% Table generated by Excel2LaTeX from sheet 'coide coverage'
\begin{table}[htbp]
  \centering
  \caption{Branch Coverage of MACdroid and the baselines}
      \resizebox{1\linewidth}{!}{
    \begin{tabular}{c|cc|r|r|r|r|r|r|r}
    \toprule
    \multirow{2}[4]{*}{\textbf{Dataset}} & \multicolumn{1}{c|}{\multirow{2}[4]{*}{\textbf{Category}}} & \multirow{2}[4]{*}{\textbf{App}} & \multicolumn{4}{c|}{\textbf{Coverage number}} & \multicolumn{3}{c}{\textbf{Coverage capability}} \\
\cmidrule{4-10}          & \multicolumn{1}{c|}{} &       & \multicolumn{1}{l|}{\textbf{GT.}} & \multicolumn{1}{l|}{\textbf{MAC.}} & \multicolumn{1}{l|}{\textbf{TEM.}} & \multicolumn{1}{l|}{\textbf{Auto.}} & \multicolumn{1}{l|}{\textbf{MAC.}} & \multicolumn{1}{l|}{\textbf{TEM.}} & \multicolumn{1}{l}{\textbf{Auto.}} \\
    \midrule
    \multirow{8}[6]{*}{FrUITeR} & \multicolumn{1}{c|}{\multirow{4}[2]{*}{News}} & \multicolumn{1}{l|}{ABC News} & 21775 & 24498 & 14072 & 16725 & 83\%  & 60\%  & 53\% \\
          & \multicolumn{1}{c|}{} & \multicolumn{1}{l|}{BuzzFeed} & 22146 & 19299 & 29640 & 19202 & 79\%  & 84\%  & 78\% \\
          & \multicolumn{1}{c|}{} & \multicolumn{1}{l|}{Fox News} & 27120 & 26496 & 13309 & 20521 & 94\%  & 47\%  & 51\% \\
          & \multicolumn{1}{c|}{} & \multicolumn{1}{l|}{Reuters News} & 10272 & 10508 & 10219 & 16492 & 99\%  & 82\%  & 82\% \\
\cmidrule{2-10}          & \multicolumn{1}{c|}{\multirow{3}[2]{*}{Shopping}} & \multicolumn{1}{l|}{Etsy} & 12561 & 11896 & 7249  & 8621  & 94\%  & 54\%  & 64\% \\
          & \multicolumn{1}{c|}{} & \multicolumn{1}{l|}{Geek} & 13860 & 12273 & 21481 & 10654 & 86\%  & 77\%  & 76\% \\
          & \multicolumn{1}{c|}{} & \multicolumn{1}{l|}{Wish} & 9055  & 8969  & 8903  & 8903  & 94\%  & 93\%  & 93\% \\
\cmidrule{2-10}          & \multicolumn{2}{c|}{\textbf{Average}} & 16684  & 16277  & 14982  & 14445  & \textbf{89\%} & \textbf{67\%} & \textbf{67\%} \\
    \midrule
    \multirow{20}[12]{*}{Lin} & \multicolumn{1}{c|}{\multirow{4}[2]{*}{Browser}} & \multicolumn{1}{l|}{Lightning} & 14294 & 13962 & 5344  & 5061  & 97\%  & 37\%  & 35\% \\
          & \multicolumn{1}{c|}{} & \multicolumn{1}{l|}{Privacy Browser} & 3400  & 3447  & 3378  & 2749  & 96\%  & 99\%  & 81\% \\
          & \multicolumn{1}{c|}{} & \multicolumn{1}{l|}{FOSS Browser} & 1432  & 1432  & 1432  & 827   & 100\% & 100\% & 58\% \\
          & \multicolumn{1}{c|}{} & \multicolumn{1}{l|}{Firefox} & 13217 & 13157 & 13324 & 2186  & 99\%  & 98\%  & 16\% \\
\cmidrule{2-10}          & \multicolumn{1}{c|}{\multirow{5}[2]{*}{To-Do}} & \multicolumn{1}{l|}{Minimal} & 2554  & 11539 & 2501  & 1543  & 98\%  & 98\%  & 49\% \\
          & \multicolumn{1}{c|}{} & \multicolumn{1}{l|}{Clear List} & 1723  & 1724  & 1732  & 1703  & 100\% & 100\% & 99\% \\
          & \multicolumn{1}{c|}{} & \multicolumn{1}{l|}{To-Do} & 3211  & 11856 & 2631  & 3376  & 91\%  & 82\%  & 91\% \\
          & \multicolumn{1}{c|}{} & \multicolumn{1}{l|}{Simply Do} & 50    & 50    & 36    & 42    & 100\% & 72\%  & 80\% \\
          & \multicolumn{1}{c|}{} & \multicolumn{1}{l|}{Shopping List} & 3144  & 10944 & 2136  & 2688  & 88\%  & 62\%  & 58\% \\
\cmidrule{2-10}          & \multicolumn{1}{c|}{\multirow{4}[2]{*}{Shopping}} & \multicolumn{1}{l|}{Geek} & 9897  & 7236  & 6979  & 6917  & 71\%  & 68\%  & 66\% \\
          & \multicolumn{1}{c|}{} & \multicolumn{1}{l|}{Yelp} & 18291 & 17265 & 10634 & 10591 & 92\%  & 57\%  & 57\% \\
          & \multicolumn{1}{c|}{} & \multicolumn{1}{l|}{Etsy} & 10663 & 11238 & 9009  & 9366  & 98\%  & 82\%  & 84\% \\
          & \multicolumn{1}{c|}{} & \multicolumn{1}{l|}{Wish} & 9053  & 17854 & 5049  & 5130  & 98\%  & 56\%  & 56\% \\
\cmidrule{2-10}          & \multicolumn{1}{c|}{\multirow{2}[2]{*}{Mail}} & \multicolumn{1}{l|}{K-9 Mail} & 3355  & 3216  & 3355  & 1433  & 95\%  & 100\% & 43\% \\
          & \multicolumn{1}{c|}{} & \multicolumn{1}{l|}{Fast Email} & 2584  & 2586  & 2584  & 1592  & 99\%  & 100\% & 60\% \\
\cmidrule{2-10}          & \multicolumn{1}{c|}{\multirow{4}[2]{*}{Calculator}} & \multicolumn{1}{l|}{Tip Calculator} & 212   & 212   & 190   & 190   & 100\% & 89\%  & 89\% \\
          & \multicolumn{1}{c|}{} & \multicolumn{1}{l|}{Simple Tip} & 1773  & 1626  & 1625  & 1810  & 92\%  & 92\%  & 100\% \\
          & \multicolumn{1}{c|}{} & \multicolumn{1}{l|}{Tip Plus} & 2619  & 2598  & 2438  & 2601  & 99\%  & 93\%  & 99\% \\
          & \multicolumn{1}{c|}{} & \multicolumn{1}{l|}{Free Tip} & 1521  & 1450  & 1412  & 1507  & 95\%  & 93\%  & 99\% \\
\cmidrule{2-10}          & \multicolumn{2}{c|}{\textbf{Average}} & 5421  & 7021  & 3989  & 3227  & \textbf{94\%} & \textbf{72\%} & \textbf{57\%} \\
    \bottomrule
    \end{tabular}}%
  \label{tab:code-coverage}%
\end{table}%

\textbf{Branch coverage of MACdroid and the baselines.} Branch coverage~\cite{gupta2000generating, grano2019branch,panichella2015reformulating} is a commonly-used code coverage metric. Compared to method coverage or block coverage, branch coverage imposes a higher standard and requires higher qualities for test cases.

In \secref{sec:evaluation}, we evaluate the effectiveness of MACdroid, TEMdroid, and AutoDroid from three perspectives. These perspectives test whether the test cases are fully executable test cases (i.e., executable-rate); whether the test cases successfully test the target functionality (i.e., success-rate); and whether the test cases align with the ground-truth test cases (i.e., perfect-rate). 
To provide a comprehensive understanding of these approaches, we also discuss the branch coverage of the test cases generated by MACdroid, TEMdroid, and AutoDroid compared with that of the ground-truth test cases.
Specifically, we design a new metric, \emph{coverage-capability}. For a given app, this metric is calculated as the ratio of common covered branches ($Cover_{common}$) between the generated test cases and the ground-truth test cases, to the total covered branches ($Cover_{gt}$) by the ground-truth test cases. We use the WALLMANUER~\cite{auer2024wallmauer} tool to calculate the branch coverage for each app.

Note that, coverage-capability and the metrics in \secref{sec:evaluation} are evaluated from different perspective. In this way, the result trends of MACdroid, TEMdroid, and AutoDroid presented here and in \secref{sec:evaluation} may be related but different.

\begin{equation}
\textit{Coverage-capability} = \textit{Cover}_\textit{common}\,/\,\textit{Cover}_\textit{gt}
\end{equation}

\tabref{tab:code-coverage} shows the branch coverage results for MACdroid (denoted as MAC.), TEMdroid (denoted as TEM.), and AutoDroid (denoted as Auto.) across all successfully \emph{instrumented} apps from the Lin and FrUITeR datasets. For each approach, we report the number of covered branches and the coverage-capability. We also report the number of branches covered by the ground-truth (denoted as GT.) test cases for reference.

Specifically, on the FrUITeR dataset, 89\% of the branches covered by the ground-truth test cases can be covered by MACdroid, while TEMdroid and AutoDroid can only cover 67\% and 67\%, respectively. The results on the Lin dataset are similar, with MACdroid covering 94\% of the branches covered by the ground-truth test cases, while TEMdroid and AutoDroid only cover 72\% and 57\%, respectively.
These results indicate that MACdroid achieves the closest match to the ground-truth in terms of branch coverage, outperforming TEMdroid and AutoDroid. Note that, the FrUITeR and Lin datasets used in this study are among the most widely-used datasets in test case migration. Both datasets include industry-level apps (e.g., ABC News~\cite{abc-news} and Firefox~\cite{firefox}), making them effectively evaluate test case migration approaches in real-world environments.

\section{Related Work}\label{sec:related}
\textbf{GUI test migration.} 
Several approaches~\cite{hu2018appflow, behrang2019test, lin2019test, liu2023test, mariani2021evolutionary, zhang2024learning} have been proposed for migrating GUI test cases between different apps.
AppFlow~\cite{hu2018appflow} utilizes a trained multi-classifier to identify widget labels, considering widgets with the same widget labels as mapped widgets. 
ATM~\cite{behrang2019test}, Craftdroid~\cite{lin2019test}, TRASM~\cite{liu2023test}, and Adaptdroid~\cite{mariani2021evolutionary} leverage different word embeddings~\cite{mikolov2013distributed, kusner2015word} to represent words in widgets and employ a manually defined matching function to map widgets. 
TEMdroid~\cite{zhang2024learning} is the first approach that trains a matching model for widget mapping.
After mapping widgets, these migration approaches generate the corresponding events and assertions for the target apps.
MigratePro~\cite{zhang2024synthesis} is a related approach for GUI test migration that seeks to improve migration techniques through test case synthesis.

The key difference between existing migration approaches and \toolNameSmall{} lies in the distinct paradigms they follow for GUI test migration. Existing migration approaches follow the widget-mapping paradigm, which maps widgets from source test cases to target test cases. In contrast, \toolNameSmall{} is based on the abstraction-concretization paradigm, which first abstracts the general test logics for the target functionalities and then uses these logics to guide the LLM to concretize the target test cases. 
Compared with existing migration approaches, the test cases generated by \toolNameSmall{} show significant improvements (see \secref{sec:rq1}).

\textbf{Functional GUI test generation.} 
Several approaches~\cite{wen2023droidbot, li2020mapping, koroglu2021functional, wen2023empowering,taeb2023axnav, jiang2023iluvui,gao2023assistgui} aim to generate GUI test cases for apps in different systems.
Among them, four approaches~\cite{wen2023droidbot, li2020mapping, koroglu2021functional, wen2023empowering} target the Android system. Li et al.~\cite{li2020mapping} input manually crafted test logics and manually selected candidate app screenshots for the target functionality and utilized a matching model to select events appearing in the screenshots. However, this approach only supports one action (i.e., click), limiting its applicability in real-world scenarios. FARLEAD-Android~\cite{koroglu2021functional} requires users to provide formal specifications as inputs, which poses a significant challenge for user adoption. 
DroidBot-GPT~\cite{wen2023droidbot} utilizes LLMs to select events based on manually crafted test logics. AutoDroid~\cite{wen2023empowering}, an advanced version of DroidBot-GPT, includes an offline stage to understand the state relationships.
For the iOS system, AXNav~\cite{taeb2023axnav} and ILvuUI~\cite{jiang2023iluvui} input test logic and app screenshots. They further employ vision-based LLMs to select events. 
AssistGUI~\cite{gao2023assistgui} is a GUI test generation approach specifically designed for the Windows system.
Note that, the source code and tools for FARLEAD-Android, AXNav, ILvuUI, and AssistGUI are not available.

The key difference between existing GUI test generation approaches and \toolNameSmall{} lies in how the test logics for the target functionalities are obtained. These approaches rely on manually crafted test logics, making the process time-consuming and impractical for large-scale apps. In contrast, \toolNameSmall{} extracts test logics from source test cases, effectively replacing the need for manually crafted test logics.

\textbf{Bug detection for mobile apps.}
According to exploration strategies, bug detection related approaches for mobile apps can be classified into four categories, i.e., random testing approaches~\cite{monkey,machiry2013dynodroid, sun2023characterizing, sun2023property}, model-based approaches~\cite{su2017guided,baek2016automated,gu2017aimdroid,lai2019goal, li2017droidbot, su2021fully, wang2022detecting, liu2022guided, yu2024practical}, systematic testing approaches~\cite{anand2012automated,gao2018AndroidTest,mao2016sapienz}, and learning-based approaches~\cite{spieker2017reinforcement,borges2018guiding,koroglu2018qbe,li2019humanoid, liu2022nighthawk, yu2024practical-intrusive, qian2020roscript, bernal2020translating}. 
Specifically, Monkey~\cite{monkey}, a widely-used approach, employs random exploration to uncover bugs. AIMDROID~\cite{gu2017aimdroid} and Stoat~\cite{su2017guided} leverage static analysis and dynamic exploration to construct a model of the app under test, and subsequently detect bugs based on this model. SCENTEST~\cite{yu2024practical} adopts a different modeling approach by collecting extensive test reports and generating event knowledge graphs to guide its exploration. These event knowledge graphs integrate user testing information, representing a valuable effort to leverage user-provided insights for enhancing automated exploration and identifying complex bugs.
% Anand et al.,~\cite{anand2012automated} developed an algorithm and a system for bug detection based on concolic testing for mobile apps.
SynthesiSE~\cite{gao2018AndroidTest} is a concolic execution approach for Android applications. Unlike traditional methods, which rely on manually written models for the Android framework, SynthesiSE dynamically infers expressions representing Android models during execution.
RoScript~\cite{qian2020roscript} and ROBOTEST~\cite{yu2024practical-intrusive} focus on embedded systems. They use robotic arms with computer vision-based algorithms to detect bugs across embedded systems. V2S~\cite{bernal2020translating} aids bug detection by automatically translating
video recordings of Android app usages into replayable test cases.

The key distinction between \toolNameSmall{} and existing bug detection approaches lies in their different purposes. Existing approaches primarily focus on GUI exploration and bug detection but lack the capability to effectively emulate user behavior for testing individual functionalities.
Therefore, to validate user behavior during the execution of specific functionalities, industry practitioners often rely on extensive manual effort to design and execute tests for individual functionalities, ensuring the correctness of individual functionalities.
In contrast, \toolNameSmall{} generates test cases based on specific test logics for individual functionalities, enabling it to emulate real-world user interactions with each functionality more accurately.

\section{Conclusion}\label{sec:conclusion}
GUI test migration aims to produce test cases for specific functionalities of a target app. Existing migration approaches follow the widget-mapping paradigm. However, test cases produced using this paradigm are often incomplete or contain bugs, making them difficult to use directly for testing target functionalities and require additional manual modifications.
In this paper, we have proposed a new paradigm for GUI test migration (i.e., abstraction-concretization paradigm). We then proposed \toolNameSmall{}, the first approach that follows this paradigm to migrate GUI test cases.
We have evaluated the effectiveness of \toolNameSmall{} on 31 real-world apps, 34 functionalities, and 123 test cases, and compared it with two state-of-the-art approaches, TEMdroid and AutoDroid. Our experimental results demonstrate the effectiveness of \toolNameSmall{} in GUI test migration.

\section*{Acknowledgements}
% \section*{Acknowledgements}
We thank the anonymous TOSEM reviewers for their valuable feedback and the insightful comments provided by Zhiyong Zhou.
This work was supported by the National Key Research and Development Program of China under Grant No. 2023YFB4503803, the National Natural Science Foundation of China under Grant No.62372005, the Young Elite Scientists Sponsorship Program by CAST (Doctoral Student Special Plan), and the Ministry of Education, Singapore under its Academic Research Fund Tier 2 (Proposal ID: T2EP20223-0043; Project ID: MOE-000613-00). Any opinions, findings and conclusions or recommendations expressed in this material are those of the author(s) and do not reflect the views of the Ministry of Education, Singapore.

% \begin{acks}
% \end{acks}

\bibliographystyle{ACM-Reference-Format}
\bibliography{main}


\end{document}